%% file: main.tex
\documentclass[lettersize,journal]{IEEEtran}
\usepackage{wrapfig}
\usepackage[normalem]{ulem}
\usepackage{cite}
\usepackage{amsmath}
\usepackage{amsthm}
\usepackage{epsfig}
\usepackage{url}
\usepackage{booktabs}
\usepackage{multirow}
\usepackage{mathrsfs}
\usepackage{amsfonts}
\usepackage{algorithm,algorithmic}
\usepackage{paralist}
\usepackage{tabularx}
\usepackage{amssymb}
\usepackage{graphicx}
\usepackage{float}
\usepackage[caption=false,font=footnotesize]{subfig}
\usepackage{xcolor}
\usepackage{upgreek}
\usepackage{mathtools}
\usepackage{hyperref}

\begin{document}

\title{Olive Branch Learning: A Topology-Aware Federated Learning Framework for Space-Air-Ground Integrated Network}

\author{Qingze~Fang, Zhiwei Zhai,
        Shuai~Yu,~\IEEEmembership{Member,~IEEE,}
        Qiong~Wu,
        Xiaowen~Gong,~\IEEEmembership{Member,~IEEE,}
        and~Xu~Chen,~\IEEEmembership{Senior~Member,~IEEE}
\thanks{Qingze Fang, Zhiwei Zhai, Shuai Yu, Qiong Wu, and Xu Chen are with the School of Computer Science and Engineering, Sun Yat-sen University, Guangzhou, China, (E-mail:~ \{fangqz, zhaizhw3\}@mail2.sysu.edu.cn,~yushuai@mail.sysu.edu.cn, wuqiong23@mail2.sysu.edu.cn, chenxu35@mail.sysu.edu.cn). Xiaowen Gong is with the Department of Electrical and Computer Engineering, Auburn University, Auburn, AL, USA, (Email: xgong@auburn.edu).}
\thanks{Part of the result has been presented in 2021 IEEE International Conference on Space-Air-Ground Computing (SAGC)~\cite{SAGC2021}.}}

%



\maketitle

\theoremstyle{definition}
\newtheorem{definition}{Definition}
\newtheorem{theorem}{Theorem}
\newtheorem{lemma}{Lemma}
\newtheorem{assumption}{Assumption}
\newtheorem{remark}{Remark}

\input{Sections/Abstract}
\input{Sections/Introduction}
\input{Sections/Related_Work}
\input{Sections/Framework}
\input{Sections/Problem_and_Algorithm}
\input{Sections/Multi-orbit_Satellite_Network}
\input{Sections/Convergence_Analysis}
\input{Sections/Performance_Evaluation}
\input{Sections/Conclusion}



%






\bibliography{References/ref}
\bibliographystyle{References/IEEEtran}

%







\end{document}

%% file: Sections/Abstract.tex
\vspace{-10pt}
\begin{abstract}
The space-air-ground integrated network (SAGIN), one of the key technologies for next-generation mobile communication systems, can facilitate data transmission for users all over the world, especially in some remote areas where vast amounts of informative data are collected by Internet of remote things (IoRT) devices
to support various data-driven artificial intelligence (AI) services.
However, training AI models centrally with the assistance of SAGIN faces the challenges of highly constrained network topology, inefficient data transmission, and privacy issues.
To tackle these challenges, we first propose a novel topology-aware federated learning framework for the SAGIN, namely Olive Branch Learning (OBL).
Specifically, the IoRT devices in the ground layer leverage their private data to perform model training locally, while the air nodes in the air layer and the ring-structured low earth orbit (LEO) satellite constellation in the space layer are in charge of model aggregation (synchronization) at different scales.
To further enhance communication efficiency and inference performance of OBL, an efficient Communication and Non-IID-aware Air node-Satellite Assignment (CNASA) algorithm is designed by taking the data class distribution of the air nodes as well as their geographic locations into account.
Furthermore, we extend our OBL framework and CNASA algorithm to adapt to more complex multi-orbit satellite networks.
We analyze the convergence of our OBL framework and conclude that the CNASA algorithm contributes to the fast convergence of the global model.
Extensive experiments based on realistic datasets corroborate the superior performance of our algorithm over the benchmark policies.

\end{abstract}


%

%% file: Sections/Introduction.tex
\section{Introduction}
%
%
%
%

\IEEEPARstart{W}{ith} the rapid development of aerospace, deep-sea exploration, and unmanned technologies, Internet of remote things (IoRT) devices have been widely adopted in remote areas (e.g., isolated islands, desert hinterlands, and the deep sea) for remote sensing and environmental monitoring~\cite{9184929}. These informative sensing data can be further utilized for artificial intelligence (AI) model training to support a large number of data-driven intelligent applications, like natural hazard analysis and vegetation extent monitoring~\cite{9420293}. However, with the absence of terrestrial communication facilities such as fifth-generation (5G) base stations in remote areas, it is challenging for the dispersed IoRT devices to transmit the collected data to a centralized server for further processing and training.

The space-air-ground integrated network (SAGIN)~\cite{8368236}, serving as a complement and extension of terrestrial communication networks, offers seamless wireless access to wide geographical areas and provides a promising solution to facilitate IoRT data transmission. Specifically, SAGIN is based on a hierarchical network architecture that integrates satellite constellation in the space layer, air nodes (e.g., unmanned aerial vehicles, airships, balloons, etc.) in the air layer, and end devices (e.g., IoT devices) in the ground layer. The satellite constellation introduced in SAGIN is responsible for global coverage, while the air nodes flying at a low altitude can provide flexible ubiquitous accessibility with low propagation latency, thus supporting the data transmission of IoRT applications~\cite{zhou2021gateway, de2015satellite}. Taking advantage of the wide coverage and large communication capacity of SAGIN, many established satellite operators are deploying their SAGIN projects (e.g., Starlink, Telesat, OneWeb, etc.)~\cite{8368236}. 
To reap the benefits of the massive data from the IoRT devices, a predominant approach is to collect the scattered IoRT data to a cloud server for centralized machine learning (ML) or deep learning (DL) model training~\cite{9205981}. Nevertheless, this kind of approach
requires a central server to continuously collect the massive volume of sensing data scattered around the IoRT devices over expensive satellite networks in SAGIN scenarios, which will lead to heavy communication costs and privacy concerns~\cite{9082655}.
To cope with these problems, a privacy-preserving distributed machine learning method termed Federated Learning (FL) is proposed~\cite{pmlr-v54-mcmahan17a}. FL enables collaboratively training a shared global ML model under the coordination of a central server by aggregating locally-computed updates while keeping the sensitive data in local clients (e.g., IoRT devices). In addition, the communication cost depends on the customizable model size and does not scale up with the increase of local data size. Thus, FL model learning is able to guarantee a reduced communication cost considering the massive device data in SAGIN scenarios.

Leveraging the privacy-preserving collaborative learning power of FL, we advocate a novel topology-aware framework, namely Olive Branch Learning (OBL), tailored to the inherent characteristics of SAGIN such as system hierarchy and specific topologies of different layers. 
Nevertheless, such SAGIN based federated learning framework still faces the following challenges. First, despite of the data heterogeneity (i.e., not independently and identically distributed data, or non-IID data) inherent in the dispersed IoRT devices, the hierarchical mechanism of OBL may introduce new dimensions of data heterogeneity (e.g., in the space layer), which may further lead to degraded model accuracy and slow convergence speed for OBL. Second, long model delivery time will occur if an air node communicates with its associated remote satellite by using several relay satellites as intermediates.
To deal with the above challenges, we devise a Communication and Non-IID-aware Air node-Satellite Assignment (CNASA) algorithm to boost the training process of OBL by considering both geographical distance and data class distribution of the air nodes into account. Furthermore, we also extend the OBL framework to the multi-orbit SAGIN and devise an efficient enhanced version of CNASA algorithm for the air node-satellite assignment in the complicated multi-orbit SAGIN network.

The main contributions of this paper are as follows: 

\begin{itemize}
\item We promote a novel topology-aware federated learning framework, namely OBL, tailored to the inherent characteristics of the SAGIN with one low earth orbit as its space layer.
\item To alleviate the negative effects of non-IID data on IoRT devices as well as the expensive time cost of model delivery, we devise the CNASA algorithm for OBL, which strikes a nice balance between the global model accuracy and the overall time cost.
\item We extend our OBL framework and CNASA algorithm to adapt to the SAGIN with a multi-orbit satellite network. Specifically, we first investigate the topological properties of the multi-orbit satellite constellation, and then redesign the inter-satellite model synchronization mechanism and CNASA algorithm to fit the specific multi-orbit SAGIN scenarios.
\item Rigorous theoretical analysis of the convergence of OBL is provided. Empirically, evaluation results based on realistic datasets demonstrate that OBL is superior to the other methods in terms of model accuracy and time cost, in SAGIN with either one orbit or multiple orbits.
\end{itemize}

The rest of this paper is organized as follows.
In Section~\ref{sec:Related_Work}, we introduce the related works most relevant to this paper. Section~\ref{sec:Framework} illustrates the OBL framework and elaborates the training process of OBL. 
The problem definition and our proposed CNASA algorithm is expounded in Section~\ref{sec:Problem_Algorithm}.
We extend our OBL framework to adapt to the multi-orbit satellite network in Section~\ref{sec:Multi-orbit}.
Section~\ref{sec:Convergence_Analysis} gives a convergence analysis of OBL.
The performance evaluation is conducted in Section~\ref{sec:Performance_Evaluation}.
At last, Section~\ref{sec:Conclusion} concludes this paper.

%% file: Sections/Related_Work.tex
\section{Related Work}
{\label{sec:Related_Work}}

Due to the global-area coverage of the space-air-ground integrated network (SAGIN), data scattered across ubiquitous end devices can be brought together to support data-driven artificial intelligence~\cite{9420293}.
Nei et al.~\cite{8612450} propose a deep learning-based method to improve the traffic control performance of SAGIN considering the satellite traffic balance.
Cheng et al.~\cite{8672604} present a deep reinforcement learning-based computing offloading approach considering the network dynamics as well as the energy and computation constraints.
Tan et al.~\cite{9592717} propose a multi-modal emotion recognition model for autonomous vehicles based on a 5G-enabled SAGIN.
These works achieve great performance in terms of network throughput or model accuracy, but the data of devices are exposed to the server, which may violate the user's privacy.

As an emerging distributed learning paradigm, federated learning (FL) preserves privacy by leaving the training process to the users~\cite{pmlr-v54-mcmahan17a}.
FL has achieved great success in many applications such as the next-word prediction on mobile phones~\cite{DBLP:journals/corr/abs-1811-03604}.
As the number of devices increases, the server suffer from excessive communication overhead inevitably, thereby deteriorating the learning performance.
To address this issue, Liu et al.~\cite{9148862} propose the client-edge-cloud hierarchical FL system where edge servers perform partial model aggregation to reduce the server's communication overhead.
However, no matter in standard FL~\cite{pmlr-v54-mcmahan17a} or hierarchical FL~\cite{9148862}, the model performance would degrade severely when the local datasets are highly imbalanced (widely known as not independently and identically distributed, non-IID).
To cope with this challenge, Mhaisen et al.~\cite{9337204} propose a heuristic-based approach to facilitate the convergence of the global model by greedily equalizing the class distribution among edge devices.
Nonetheless, the authors ignore the communication delay within the client-edge network, which in some cases can greatly increase the overall delay.
Other promising research directions for hierarchical FL include the resource allocation~\cite{9562560, 9479786}.


Nevertheless, the above FL approaches do not take into account the unique characteristics of SAGIN such as the network topology properties, and directly adopting such FL schemes would lead to poor performance.
Thus, how to efficiently deploy FL into SAGIN remains to be carefully considered.
Existing works on the deployment of FL focus on the air and ground segments of SAGIN.
For example, Liu et al.~\cite{9184079} propose an FL-based air quality sensing framework in the aerial-ground integrated network, where unmanned aerial vehicles (UAVs) collaboratively monitor the air quality index (AQI) 
without sharing the raw data.
Zhang et al.~\cite{9210077} present a UAV-aided image classification approach for area exploration.
In their scheme, FL is considered for training collected images in each UAV locally, and a ground server is in charge of model aggregation.
Dong et al.~\cite{9520337} propose a novel architecture called UaaIS (UAVs as an intelligent service) for the air-ground integrated network and verify the efficiency of UaaIS in the scenario of UAV-enabled FL.
All of these works focus on the deployment of FL in the air-ground integrated network but pay no attention to the space layer, an essential component of SAGIN.
Instead, we attach importance to the space layer of SAGIN and leverage the Ring Allreduce algorithm~\cite{allreduce} for fast model aggregation to make full use of the ring topology of the satellite orbit.

%% file: Sections/Framework.tex
\section{Olive Branch Learning Framework}
{\label{sec:Framework}}

\subsection{Model Training Scenario in SAGIN}
\begin{figure}[t]
\includegraphics[width=0.49\textwidth]{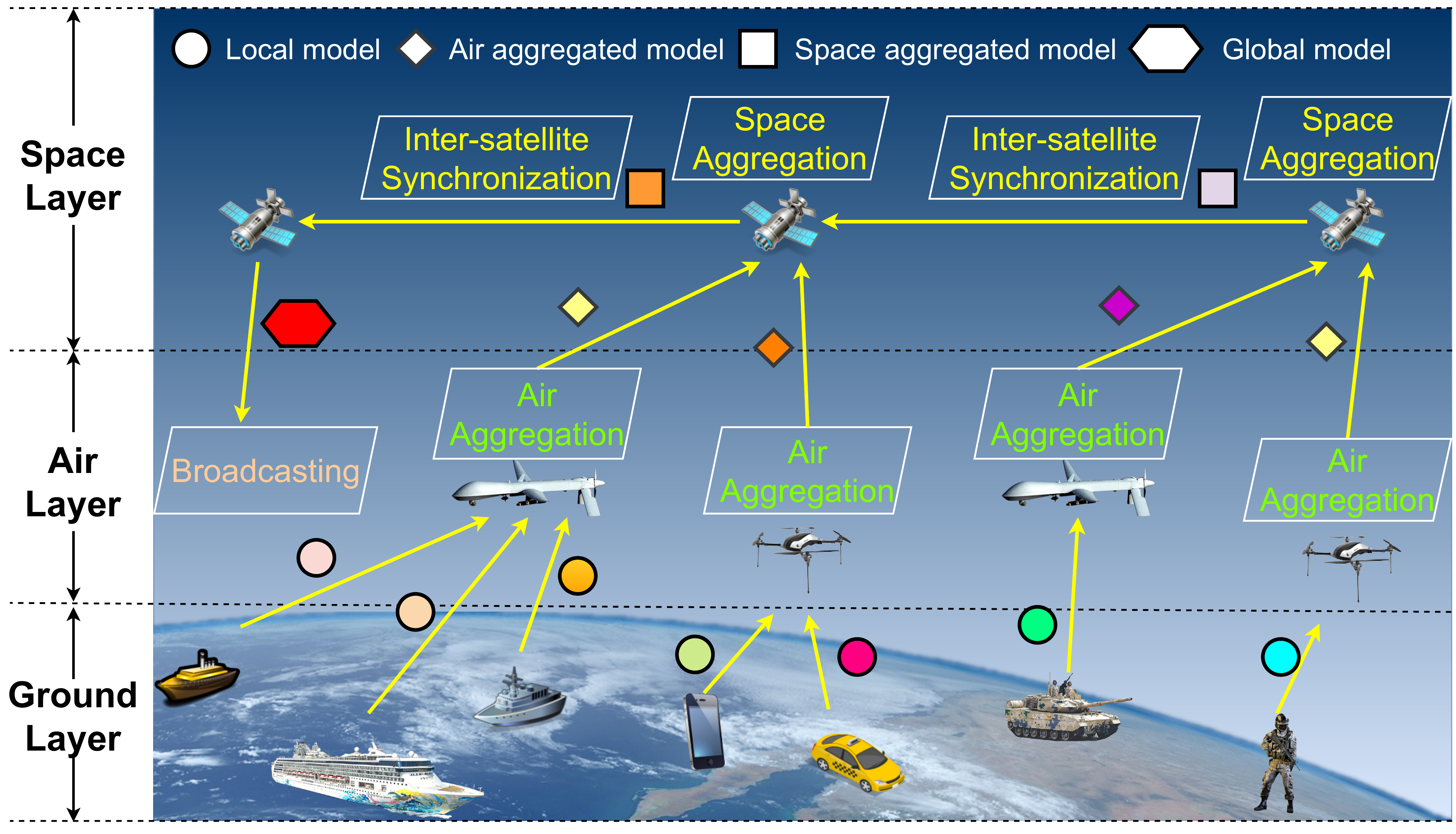}
\caption{AI model training based on federated learning in the space-air-ground integrated network.}
\label{fig:FL-SAGIN}
\end{figure}

We consider a collaborative model training scenario in the SAGIN, which naturally possesses a hierarchical architecture with three layers: space layer, air layer, and ground layer. As depicted in Fig.~\ref{fig:FL-SAGIN}, the IoRT devices in the ground layer conduct environmental monitoring and generate massive information-rich data for AI model training.
The air nodes in the air layer hover over the equator to provide flexible ubiquitous accessibility with the satellites in the space layer for the IoRT devices in their coverage areas, which can be classified into two different types, that is, high-altitude platforms (HAPs) (e.g., airships and balloons) that operate at the stratospheric altitudes in typically quasi-stationary locations~\cite{9583591, 9174937} and low-altitude platforms (LAPs) (e.g., helicopter and multicopter) that can also stay stationary in the air~\cite{8869712}. For the air node (e.g., UAV) which moves within the coverage area of its associated satellite (which is usually very large), the connectivity and link quality between the air node and its associated satellite or IoRT devices do not change with its mobility.
The satellites are evenly distributed in the equatorial low earth orbit to provide wide coverage.
To benefit from the massive data from the IoRT devices, the centralized model training requires IoRT devices to upload raw data to a cloud server via SAGIN, but this leads to data privacy leakage risks and significant communication costs. To address these issues, we propose to leverage the popular federated learning to train a global shared model for SAGIN without compromising data privacy.

\subsection{Framework Overview of Olive Branch Learning}


As shown in Fig.~\ref{fig:FL-SAGIN}, each IoRT device in the ground layer trains its own model with local data samples, while nodes in the air layer and space layer are in charge of model aggregation with the models received from their lower layers. After conducting satellite model aggregation by aggregating model updates sent from its associated air nodes, each satellite maintains a satellite model that contains the information from its associated IoRT devices\footnote{We also allow the case that some powerful IoRT device not covered by any air nodes can send its local model directly to the satellite.}. As there are many satellites in the space layer of SAGIN, inter-satellite model synchronization is further needed to obtain a global model for the whole SAGIN.

\begin{figure}[t]
\centering
\includegraphics[scale=0.262]{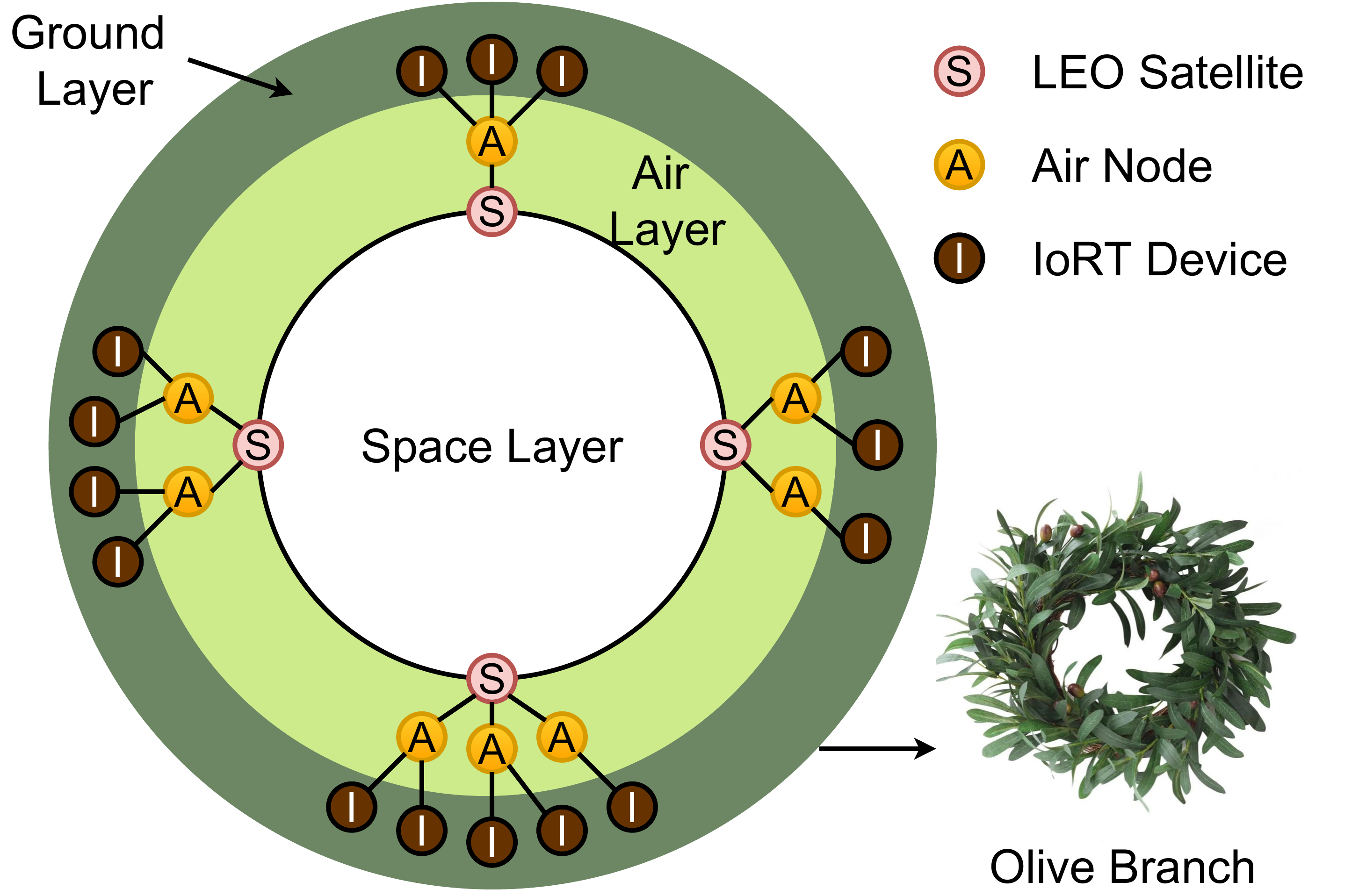}
\caption{Structure overview of Olive Branch Learning.}
\label{fig:Olive_Branch_Learning}
\end{figure}

With the inspirations above, we advocate a novel privacy-preserving learning framework named Olive Branch Learning (OBL), due to its similar structure with the olive branch, as shown in Fig.~\ref{fig:Olive_Branch_Learning}. OBL well captures the hierarchy and specific topologies of different layers in SAGIN. Specifically, IoRT devices in the ground layer are prone to send model updates to the satellites in the space layer with the air nodes in the air layer serving as intermediates. Hence, the ground-air-space communication naturally forms a star architecture with the satellite acting as the central node (aggregator). Accordingly, we can integrate the standard synchronous federated learning (e.g., the canonical FedAvg approach~\cite{pmlr-v54-mcmahan17a}) for satellite model aggregation. For inter-satellite model synchronization in the same orbit, a common way is to reform these satellites into a star structure by assigning one satellite as the master to coordinate model aggregation~\cite{9417459}, which leads to an increasing multi-hop communication cost within the space layer. Differently, the proposed OBL framework leverages the Ring Allreduce algorithm~\cite{allreduce}, which fully unleashes the benefits of the ring topology formed by the satellites in the same orbit, for fast and efficient additive model aggregation among satellites.

We should emphasize that the movement of satellites does not impact the learning performance of OBL from the perspective of IoRT devices. Following the virtual node strategy in~\cite{8932318}, we can assume that there are many fixed logical satellite positions with the same number of satellites, with each logical satellite position filled by the nearest satellite. Thus, the topology including the logical satellite positions, the air nodes, and the IoRT devices is fixed.
The above assumption naturally holds in SAGIN due to the following facts:
1) in reality, the time needed for satellite model aggregation through ground-air-space communication can be negligible for the slow-moving satellite to leave the current position; 
2) when the satellite of each logical satellite position is switched, all air nodes only need to establish links with the newly arrived satellites and send the models to the satellites for aggregation, which does not change the air nodes corresponding to each logical satellite position.
Therefore, the exact position of the satellites does not matter. From a long-term perspective, the IoRT devices are loosely coupled with the satellites, meaning that the IoRT devices can be dynamically assigned to the nearest satellite according to the movements of satellites.

Nevertheless, the hierarchy 
of OBL, together with the non-IID data distribution among IoRT devices, may introduce new dimensions of data heterogeneity (e.g., in the space layer), leading to degraded global model performance. This negative impact can be dampened by carefully deciding the assignments between air nodes and satellites, which will be elaborated in Section~\ref{sec:Problem_Algorithm}.



\subsection{Training Process of Olive Branch Learning}
Assume that there are $N_S$ satellites in the SAGIN network, and each satellite is in charge of the model aggregation among a non-overlapping set of IoRT devices $\{\mathcal{N}_k\}_{k=1, ..., N_S}$  with the associated air nodes serving as intermediates. Thus, the set of all IoRT devices scattered in the ground layer can be denoted as $\mathcal{N} = \cup_{1 \le k \le N_S} \mathcal{N}_k$. The local dataset of IoRT device $i$ in $\mathcal{N}_k$ is represented as $\mathcal{D}_{k,i}$, and $\mathcal{D}_k=\cup_{i \in \mathcal{N}_k} \mathcal{D}_{k,i}$. As a result, the objective of OBL is to minimize the loss function over all the training data $\mathcal{D}=\cup_{1 \le k \le N_S}\mathcal{D}_k$, i.e.,
\begin{align}
\small
\label{eq:obj_Olive-Branch-Learning}
    \min_{\boldsymbol w}{ F(\boldsymbol w) } &= \sum_{1 \le k \le N_S} \frac{|\mathcal{D}_k|}{|\mathcal{D}|} F_{k}(\boldsymbol w)\notag\\
    &= \sum_{1 \le k \le N_S} \frac{|\mathcal{D}_k|}{|\mathcal{D}|} \sum_{ i \in \mathcal{N}_k }{ \frac{ |\mathcal{D}_{k, i}| }{ |\mathcal{D}_k| } F_{k, i}(\boldsymbol w) },
\end{align}
where $F(\boldsymbol w)$ can be rephrased as a linear combination of the empirical objectives $F_k(\boldsymbol w)$ over dataset $\mathcal{D}_k$ for satellite $k \in \{1,2,...,N_S\}$. $F_{k, i}(\boldsymbol w)$ is the local loss function over $\mathcal{D}_{k,i}$ for IoRT device $i$ that associated with satellite $k$ through air nodes. 

Tailored to the topology of different layers in SAGIN, OBL devises satellite model aggregation and inter-satellite synchronization to train a global model in an iterative fashion, with each iteration called a global training round. The two key components for the training process of OBL are elaborated as below.

\textbf{1) Satellite model aggregation.} As the ground-air-space topology naturally forms the star structure, we integrate the standard federated learning (e.g., FedAvg~\cite{pmlr-v54-mcmahan17a}) approach for satellite model aggregation.
To reduce the ground-air-space communication cost, each IoRT device performs multiple rounds of local model training within each satellite model aggregation.
Specifically, in each local training round $t$, all the IoRT devices associated with satellite $k$ train their models $\boldsymbol w^{(t)}_{k,i}$ locally, as expressed in
\begin{equation}
\small
    \boldsymbol w^{(t)}_{k,i} = \boldsymbol w^{(t-1)}_{k,i} - \eta \nabla F_{k,i}(\boldsymbol w^{(t-1)}_{k,i}).
\end{equation}
After $\tau_1$ rounds of local model training (i.e., $t \text{ mod } \tau_1 = 0$), each satellite collects the 
models from its associated IoRT devices to obtain an aggregated satellite model $\boldsymbol w^{(t)}_{k}$,
as expressed in  
\begin{equation}
\small
    \boldsymbol w^{(t)}_{k} = \sum_{ i \in \mathcal{N}_{k} }{ \frac{ |\mathcal{D}_{k,i}| }{ |\mathcal{D}_{k}| } \boldsymbol w^{(t)}_{k,i} }.
\end{equation}
To reduce the communication cost, we also allow that partial local model aggregation on the air nodes can be conducted before delivering the models to the associated satellites.
After the satellite model aggregation, each satellite will broadcast the updated satellite model to its associated IoRT devices (e.g., via using the air nodes as relays) for the next round of local training.

\textbf{2) Inter-Satellite model synchronization.} After several satellite model aggregations, all  satellites in the same orbit cooperate to perform inter-satellite model synchronization,
as in 
\begin{equation}
\small
    \boldsymbol w^{(t)} = \sum_{ 1 \le k \le N_S }{ \frac{ |\mathcal{D}_{k}| }{ |\mathcal{D}| }\boldsymbol w^{(t)}_{k} },
\end{equation}
where $\boldsymbol w^{(t)}$ denotes the global model in local training round $t$. It is worth noting that the inter-satellite model synchronization can be conducted every $\tau_2$ times (i.e., $t \text{ mod } \tau_1\tau_2 = 0$) of satellite model aggregation in consideration of the non-negligible inter-satellite communication cost. After the inter-satellite model synchronization, each satellite will broadcast the updated global model to its associated IoRT devices for the next round of global training.
The number of global training rounds can be preset before training starts or controlled by the early stopping strategy~\cite{Yao2007}.

The feasibility of our OBL framework in the SAGIN scenarios can be fully guaranteed. Firstly, the hierarchical federated learning architecture can reduce the communication overhead, and delivering models instead of the raw data can help reduce the communication cost, which is compatible with the communication and storing capability of links in SAGIN. Moreover, model aggregation performed on the air nodes and satellites has low computing complexity and can be well adapted to the computing capability of the air nodes and satellites. Besides, model compression techniques, including quantization and sparsification~\cite{DBLP:journals/corr/KonecnyMYRSB16}, can be easily adapted to our OBL framework to further reduce the communication, computation, and storage overhead.


To sum up, the total training process of OBL can be given formally in Definition~\ref{def:Olive-Branch-Learning}.
\begin{definition}[Olive Branch Learning]
\label{def:Olive-Branch-Learning}
The evolution of local model $\boldsymbol w^{(t)}_{k,i}$ on IoRT device $i$ in $\mathcal{N}_k$ is formalized as
\begin{equation}
\small
\label{eq:w_Olive-Branch-Learning}
    \boldsymbol w^{(t)}_{k,i} = \begin{dcases}
        \boldsymbol w^{(t-1)}_{k,i} - \eta \nabla F_{k,i}(\boldsymbol w^{(t-1)}_{k,i}) & \text{if} ~ t \text{ mod } \tau_1 \ne 0, \\
        \boldsymbol w^{(t)}_{k} & \text{if} ~ \begin{aligned}[t]
            & t \text{ mod } \tau_1 = 0 \text{ and }\\
            & t \text{ mod } \tau_1\tau_2 \ne 0,
        \end{aligned} \\
        \boldsymbol w^{(t)} & \text{if} ~ t \text{ mod } \tau_1\tau_2 = 0,
    \end{dcases}
\end{equation}
where $\boldsymbol w^{(t)}_{k} = \sum_{ i \in \mathcal{N}_{k} }{ \frac{ |\mathcal{D}_{k,i}| }{ |\mathcal{D}_{k}| }[ \boldsymbol w^{(t-1)}_{k,i} - \eta \nabla F_{k,i}(\boldsymbol w^{(t-1)}_{k,i}) } ]$
and $\boldsymbol w^{(t)} = \sum_{ 1 \le k \le N_S }{ \frac{ |\mathcal{D}_{k}| }{ |\mathcal{D}| }\boldsymbol w^{(t)}_{k} }$.
\end{definition}

To well suit the ring structure of satellites, the inter-satellite model synchronization can be realized by some decentralized alternatives to federated learning (e.g., the gossip protocol~\cite{blot2016gossip} or the Ring Allreduce Algorithm~\cite{allreduce}) that do not require a central control (e.g., a master satellite) to coordinate the synchronization, which can be triggered by the first satellite to complete the satellite model aggregation. Nevertheless, the gossip protocol requires all the satellites to frequently deliver models to their neighboring satellites such that each satellite's model can be propagated through the whole satellite network, which incurs excessive communication overhead.
While Ring Allreduce algorithm reaps the benefits of model segmentation to deliver model parameters via one-hop communications between nearby satellites\footnote{For the cases where two neighboring satellites do not possess a direct link, we can use the ground station between them as the relay.} in parallel so that all the satellites can obtain a weighted average model in a fast manner. Hence, we choose Ring Allreduce algorithm for efficient inter-satellite model synchronization.

\begin{figure*}[t!]
    \centering
    \includegraphics[width=1\linewidth]{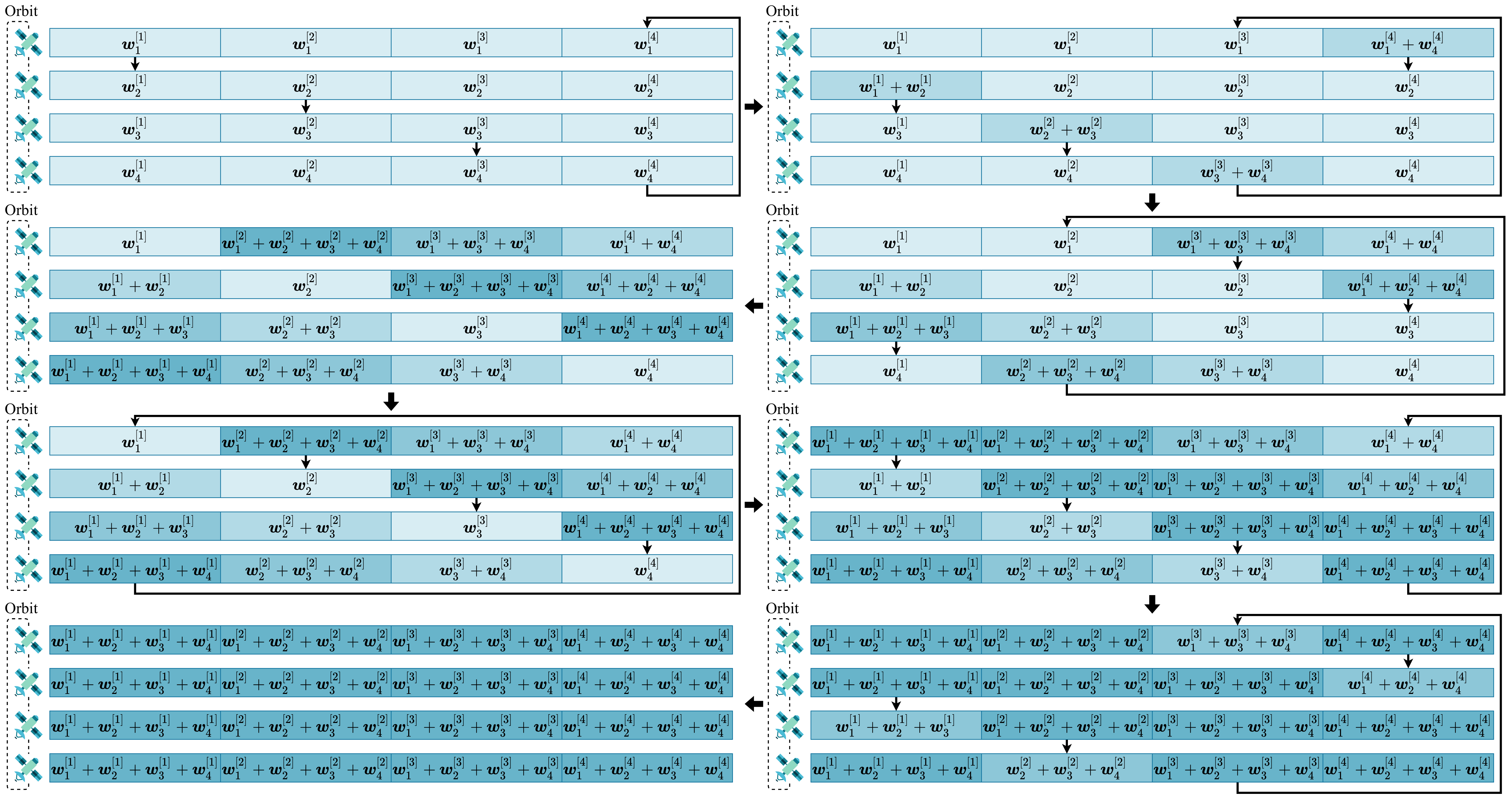}
    \vspace{-10pt}
    \caption{An illustration of the procedure of the Ring Allreduce algorithm with 4 satellites in the same orbit.}
    \label{fig:Ring_Allreduce}
    \vspace{-15pt}
\end{figure*}

The detailed procedure for inter-satellite model synchronization using Ring Allreduce algorithm is described as follows.
Fig.~\ref{fig:Ring_Allreduce} illustrates an example of the Ring Allreduce algorithm running on four satellites in one orbit.
At first, the satellites segment their models into $N_S$ chunks, where the $x$-th chunk of the satellite $k$'s model is denoted by $\boldsymbol w_k^{[x]}$ as in Fig.~\ref{fig:Ring_Allreduce}.
Then, the satellites perform $N_S - 1$ iterations of the scatter-reduce.
In the first iteration, satellite $k$ sends $\boldsymbol w_k^{[k]}$ to satellite $k + 1$ and receives $\boldsymbol w_{k - 1}^{[k - 1]}$ from satellite $k - 1$ simultaneously.
In the next iteration, satellite $k$ performs the reduction (accumulation) operation to the received chunk from satellite $k - 1$ and its own chunk and sends the reduced chunk to satellite $k + 1$.
By repeating such iteration of scatter-reduce $N_S - 1$ times, each satellite obtains a different chunk of the synchronized model, as shown in the fourth sub-graph of Fig.~\ref{fig:Ring_Allreduce}, where the color depth indicates the progress of chunk synchronization.
After that, the satellites perform $N_S - 1$ iterations of chunk sends and receives so that each satellite can acquire all the synchronized chunks, which is called the allgather.
The allgather proceeds similarly to the scatter-reduce and the only difference is that the satellites simply overwrite the received chunks instead of performing the reduction operation.
At last, each satellite stitches all chunks to obtain a global model, and the global model obtained by each satellite is the same.
During the entire process of the Ring Allreduce algorithm, each satellite sends and receives chunks $N_S - 1$ times for the scatter-reduce and $N_S - 1$ times for the allgather, and the chunk parameters size is $M/N_S$ where $M$ denotes the model parameters size.
Thus, the total size of parameters delivered to and from every satellite is $2(N_S - 1)M/N_S$, which is independent of $N_S$, revealing the efficiency of the Ring Allreduce algorithm.
However, if the gossip protocol is used, it takes at least $\log(N_S)$ gossip cycles for a satellite to send its model to all other satellites, and the inter-satellite model synchronization requires $N_S \log (N_S)$ gossip cycles. Therefore, the total size of parameters delivered to and from every satellite is $N_S \log (N_S) \cdot M$, which is much larger than that in Ring Allreduce algorithm.

%% file: Sections/Problem_and_Algorithm.tex
\section{Communication and Non-IID-aware Air node-Satellite Assignment}
{\label{sec:Problem_Algorithm}}

OBL promotes the collaborative model training for SAGIN by well capturing the hierarchy and specific topologies of the three layers in SAGIN.
However, the assignment between air nodes and satellites remains to be determined, which has a significant impact on the global model accuracy as well as the total communication time cost.
To enhance the performance of OBL in terms of these two aspects\footnote{We will also consider the energy overhead optimization in future work.}, we first formulate our optimization problem with the air node-satellite assignment as its vital controllable knob, then further propose the Communication and Non-IID-aware Air node-Satellite Assignment (CNASA) algorithm, which aims at minimizing the overall time cost of training a model and meanwhile achieve a high global model accuracy.

\subsection{Problem Definition}
{\label{sec:Problem_Definition}}

In SAGIN, the assignment between air nodes and satellites is a dominant factor for the total communication time cost due to the limited inter-satellite link bandwidth~\cite{electronics8111247}.
For example, if an air node is assigned to the satellite far away from it, it may require a huge amount of relay satellites to help upload the model, resulting in expensive communication time costs.
To clearly depict the impact of the air node-satellite assignment on the communication time cost, we investigate the overall time cost of each global training round throughout the learning process.
In each global training round, the overall time cost includes the time cost of satellite model aggregation and inter-satellite model synchronization.
Within each satellite model aggregation, the time cost can be divided into two parts, namely communication time cost and computation time cost.
The communication time cost in global training round $i$ can be expressed as
\begin{equation}
\small
T_{comm}(i) = \tau_2 \cdot ( T^{SG} + T^{GA} + T^{AS} + N^{SS}(i) \cdot T^{SS} ),
\end{equation}
where $T^{SG}$, $T^{GA}$, $T^{AS}$, and $T^{SS}$ denote the end-to-end delay of model delivery between a satellite and an IoRT device, an IoRT device and an air node, an air node and its access satellite, two neighboring satellites, respectively.
These can be different for different IoRT device-air node pairs, air node-satellite pairs, and satellite-satellite pairs, but for simplicity of notation, we omit the notations of different pairs.
$N^{SS}(i)$ denotes the maximum number of model forwarding times required by relay satellites to help the air nodes upload models to their target satellites in global training round $i$, which is related to the assignment between air nodes and satellites.

The end-to-end delay of model delivery consists of transmission delay and propagation delay.
The transmission delay $T^{XY}_{trans}$ deals with the size of model parameters $M$,
the rain attenuation ratio $\Lambda$, the channel bandwidth $B^{XY}$, the transmission power $p_{XY}$, the channel fading coefficient $h_{XY}$, and the noise $\sigma$~\cite{9328513}, which is expressed as
\begin{equation}
\small
T^{XY}_{trans} = M/\left[\Lambda B^{XY} \log_2 \left(1 + \frac{p_{XY} \cdot \left|h_{XY}\right|^2}{\sigma^2}\right)\right],
\end{equation}
where $XY \in \{SG, GA, AS, SS\}$.
The propagation delay is related to the length of physical link $L^{XY}$ and the propagation speed in wireless medium $V^{XY}$, which is expressed as
\begin{equation}
\small
T^{XY}_{prop} = \frac{L^{XY}}{V^{XY}}.
\end{equation}
Thus, the end-to-end delay of model delivery $T^{XY}$ can be represented as
\begin{equation}
\small
T^{XY} = T^{XY}_{trans} + T^{XY}_{prop}.
\end{equation}

The computation time cost within each satellite model aggregation consists of the model training time cost and the model aggregation time cost, which can be expressed as
\begin{equation}
\small
T_{comp} = \tau_2 \cdot ( \tau_1 \cdot T_{train} + T^{A}_{agg} + T^{S}_{agg} ),
\end{equation}
where $T_{train}$ denotes the model training time cost on IoRT devices.
$T^{A}_{agg}$ and $T^{S}_{agg}$ denote the model aggregation time cost on air nodes and satellites, respectively.
The model training time cost depends on the floating-point operations (FLOPs) of the model, the number of data samples $|\mathcal{D}_{ep}|$ used in one local training epoch, the number of local training epochs $N_{ep}$, and the floating-point operations per second (FLOPS) of the IoRT device for model training $FLOPS^{G}$, as in
\begin{equation}
\small
T_{train} = \frac{FLOPs \cdot |\mathcal{D}_{ep}| \cdot N_{ep}}{FLOPS^{G}}.
\end{equation}
The model aggregation time cost is related to the size of model parameters $M$, the number of received models $N^Z_M$, and the FLOPS of the aggregator node $FLOPS^{Z}$, which is expressed as
\begin{equation}
\small
T^{Z}_{agg} = \frac{M \cdot N^Z_M}{FLOPS^{Z}},
\end{equation}
where $Z \in \{A, S\}$.

Similarly, we can obtain the time cost of inter-satellite model synchronization, which is related to the total size of parameters delivered to and from every satellite as well as the number of parameter deliveries.
The time cost of inter-satellite model synchronization in global training round $i$ is linear with the number of satellites $N_S$, which can be represented as
\begin{equation}
\small
T_{sync} = 2(N_S - 1) \left( \frac{T^{SS}_{trans}}{N_S} + T^{SS}_{prop} + \frac{M}{N_S \cdot FLOPS^{S}} \right).
\end{equation}
Hence, the overall time cost of the whole training process is expressed as
\begin{equation}
\small
\sum_{i}(T_{comm}(i) + T_{comp} + T_{sync} ),
\end{equation}
where the communication time cost can be reduced by assigning closer air nodes to each satellite.

The global model accuracy of OBL can also be affected by the air node-satellite assignment strategy.
Guided by the insight in~\cite{9337204}, the accuracy of the global model can be boosted by assigning air nodes with as diverse class distribution vectors as possible to each satellite.
In other words, air nodes are assigned to satellites so that the class distribution vectors of air nodes for all satellites are as similar as possible.
Here, the class distribution vector of an air node refers to the class distribution vector of the training data under air node $j$, which can be obtained by averaging the class distribution vectors of IoRT devices in its coverage area.
Specifically, the class distribution vector of the IoRT device $i$ in $\mathcal{N}_k$ is the distribution vector over labels at its dataset $\mathcal{D}_{k,i}$, which can be denoted by $\boldsymbol{p}_{k,i}$ with its L1 norm equals to one, i.e., $\Vert \boldsymbol{p}_{k,i} \Vert_1 = 1$.
Suppose that the air node $j$ is assigned to the satellite $k$, the class distribution vector of the air node $j$ can be represented as
\begin{equation}
\small
\tilde{\boldsymbol{p}}_{j} = \frac{\sum_{i \in \mathcal{N}_{k,j}} |\mathcal{D}_{k, i}| \cdot \boldsymbol{p}_{k, i}}{\sum_{i \in \mathcal{N}_{k,j}} |\mathcal{D}_{k, i}|},
\end{equation}
where $\mathcal{N}_{k,j}$ denotes the set of IoRT devices in the coverage area of air node $j$.
The class distribution information of the IoRT devices is only revealed in an aggregated format by using secure multiparty computation (e.g., privacy-preserving k-secure sum protocol~\cite{sheikh2009privacy}), hence no violation of individual class distribution privacy happens.

Hence, we aim to achieve a better trade-off between the overall time cost and the global model accuracy for efficient OBL model training by carefully designing the air node-satellite assignment mechanism.
We denote the set of air nodes as $\mathcal{A}$, the set of satellites as $\mathcal{S}$, and the air node-satellite assignment as $f: \mathcal{A} \rightarrow \mathcal{S}$. The objective function is given as
\begin{equation}
\small
\label{eq:objective_function}
\min_{f} (1-\gamma) \sum_{i}(T_{comm}(i) + T_{comp} + T_{sync} ) - \gamma acc,
\end{equation}
where $acc$ denotes the accuracy of the global model that can be improved by assigning the air nodes with the most different class distribution vectors to each satellite, and $\gamma$ denotes the weighting factor balancing between overall time cost and global model accuracy.
Indeed, $T_{comp}$ and $T_{sync}$ are included to make our system model more complete, while $T_{comm}(i)$ is the dominating part of the overall time cost to be optimized. However, the air node-satellite assignment may also impact the computation time of the satellite model aggregation (i.e., sum averaging operation, although which is very insignificant compared to the communication time cost), since the number of air nodes associated with a satellite can be different.

\begin{figure}[t!]
\centering
\includegraphics[width=0.49\textwidth]{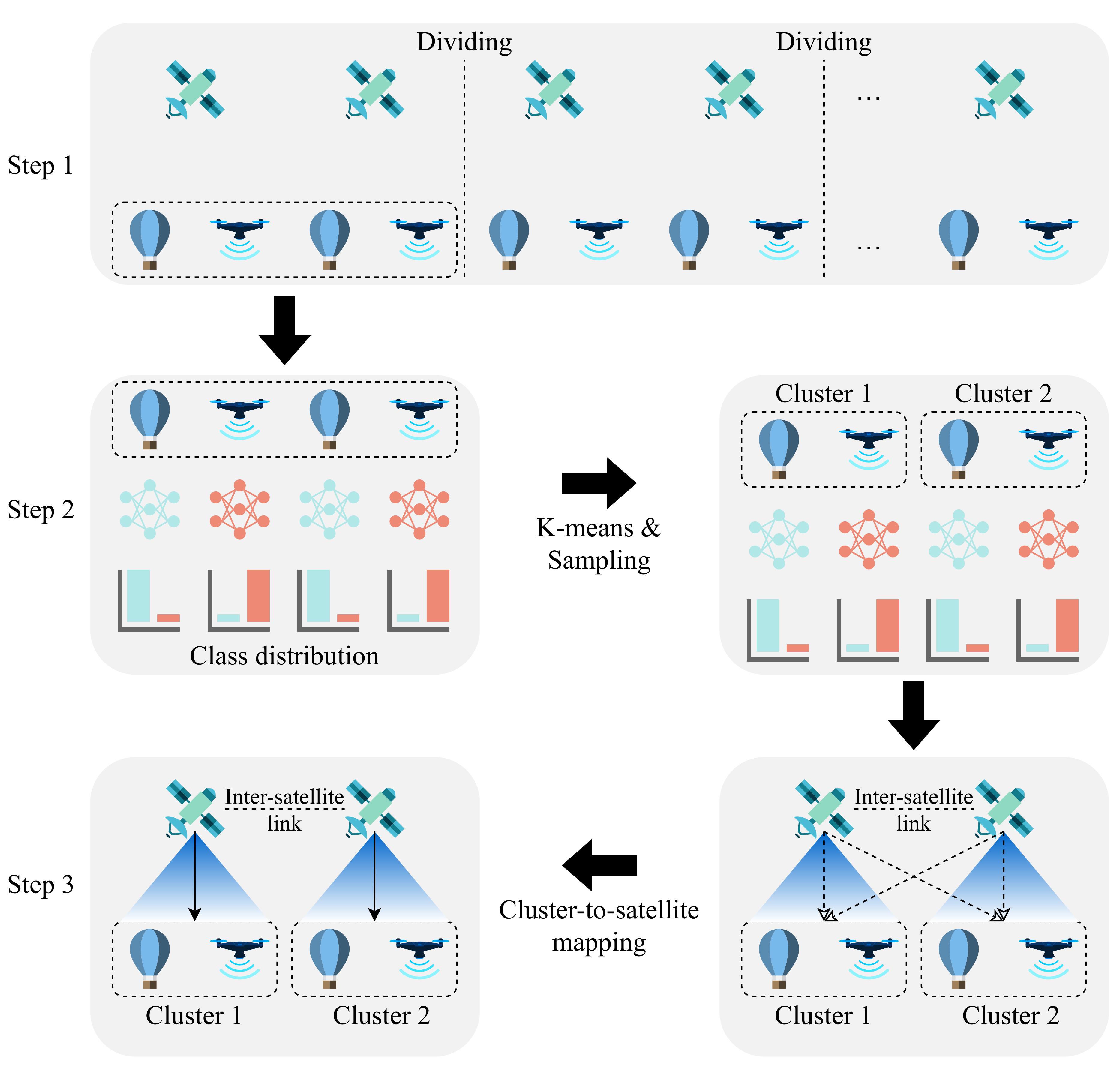}
\vspace{-10pt}
\caption{Schematic diagram of the CNASA algorithm.}
\label{fig:Clustering_and_Mapping}
\vspace{-15pt}
\end{figure}

\subsection{Algorithm Design}
{\label{sec:Algorithm_Design}}

The optimization problem defined above has a large decision space, i.e. $N_S$ to the power of $N_A$, where $N_A$ denotes the number of air nodes.
Finding the optimal solution in such a large decision space is impractical, especially when the number of both satellites and air nodes is large.
A simple heuristic-based method \cite{9337204} could be that assigning the air nodes with the most different class distribution vectors to each satellite neglecting the geographical distance between air nodes and satellites.
However, the model uploading from air nodes to satellites may become a dominant communication bottleneck, and the number of model forwarding times $N^{SS}$ may become huge, thereby increasing the overall time cost.
To strike a nice balance between communication time cost and global model accuracy, we propose a novel algorithm called Communication and Non-IID-aware Air node-Satellite Assignment (CNASA).
Specifically, we first divide the satellites and air nodes into several isolated partitions based on geographical locations to ensure that the model uploading from air nodes to satellites within each partition is communication-efficient, then assign the air nodes with diverse class distribution vectors to each satellite using a clustering-based approach.
The CNASA algorithm is expounded as follows.



\begin{enumerate}
\item \label{alg:Step_1} \emph{Step 1: Geographical-distance-based division.}
To limit the communication time cost of model uploading from air nodes to satellites, the satellites are divided equally into several non-overlapping partitions based on their geographical locations, and each partition includes $N_{geo}$ satellites, which are continuous in orbit.
Then, the air nodes can also be divided into several partitions $\mathcal P=\{\mathcal P_1, \mathcal P_2, ..., \mathcal P_{N_S / N_{geo}}\}$ according to the division relationship of their closest access satellites, with partition $\mathcal P_i$ including $ N^{(i)}_A $ air nodes.
\item \label{alg:Step_2} \emph{Step 2: Class-distribution-based clustering.} After dividing the air nodes into $N_S / N_{geo}$ partitions, the geographic distance between air nodes in each partition is limited by the boundary of the partition.
To boost global model accuracy, we group the air nodes within each partition into $N_{geo}$ clusters so that the class distribution vectors of all clusters are as similar as possible.
Specifically, for each partition, we first use the k-means clustering algorithm to divide the air nodes into $N^{(i)}_A/N_{geo}$ homogeneous groups (i.e., the class distribution vectors of air nodes within each group are similar) according to their class distribution vectors and initialize $N_{geo}$ empty clusters.
For each cluster, we sample one air node from each homogeneous group without replacement and add it to the cluster.
If a homogeneous group becomes empty, we randomly sample an air node from the other homogeneous groups.
Eventually, we obtain $N_{geo}$ clusters for each partition and each cluster contains $N^{(i)}_A/N_{geo}$ air nodes with large differences in class distribution to achieve high global model accuracy.
\item \label{alg:Step_3} \emph{Step 3: Cluster-to-satellite mapping.} Since we divide air nodes into several partitions in Step~\ref{alg:Step_1}, each cluster can only select the satellites in the partition it belongs to.
Each partition can be treated as a graph, 
where clusters and satellites can be regarded as vertices in the graph, and the weight of the edge between each cluster vertex and each satellite vertex is the total model delivery time from the air nodes in each cluster to each satellite.
Our goal is to find a cluster-to-satellite mapping that minimizes the total time overhead of model uploading.
Thus, the cluster-to-satellite mapping is equivalent to the minimum weight matching in bipartite graphs.
We solve this problem using the Jonker-Volgenant algorithm~\cite{7738348}, a faster variant of the Hungarian algorithm.
\end{enumerate}

\renewcommand{\algorithmicrequire}{ \textbf{Input:}} 
\renewcommand{\algorithmicensure}{ \textbf{Output:}} 
\begin{algorithm}[t]
\small
    \caption{CNASA}
    \label{alg:Clustering_and_mapping}
    \begin{algorithmic}[1]
        \REQUIRE the set of air nodes $\mathcal A = \{A_1, A_2, ..., A_{N_A}\}$, the set of satellites $\mathcal S = \{S_1, S_2, ..., S_{N_S}\}$, the number of satellites in each geographical distance based division $N_{geo}$
        \ENSURE the mapping from air nodes to satellites $f$ 
        \STATE Divide the satellites into $N_S / N_{geo}$ partitions equally and without overlapping based on their geographical location, and accordingly, the air nodes are divided into $N_S / N_{geo}$ non-overlapping partitions $\mathcal P=\{\mathcal P_1, \mathcal P_2, ..., \mathcal P_{N_S / N_{geo}}\}$
        \FOR {each partition $\mathcal P_i$}
            \STATE Use k-means clustering algorithm to divide the air nodes into $N^{(i)}_A / N_{geo}$ homogeneous groups $\mathcal R=\{\mathcal R_1, \mathcal R_2, ..., \mathcal R_{N^{(i)}_A / N_{geo}}\}$ according to the class distribution of the air nodes
            \STATE Create $N_{geo}$ empty clusters $\mathcal C = \{\mathcal C_1, \mathcal C_2, ..., \mathcal C_{N_{geo}}\}$
            \FOR{each cluster $\mathcal C_j$}
                \FOR {each homogeneous group $\mathcal R_k$}
                    \IF {$\mathcal R_k$ is not empty}
                        \STATE Sample one air node from $\mathcal R_k$ without replacement, and add it to $\mathcal C_j$
                    \ELSE
                        \STATE Sample one air node from another non-empty group without replacement, and add it to $\mathcal C_j$
                    \ENDIF
                \ENDFOR
            \ENDFOR
            \STATE Use Jonker-Volgenant algorithm to find cluster-to-satellite mapping $f_{cs}$ with the least model delivery time
            \STATE Extract the air node to satellite mapping from $f_{cs}$ and merge it into $f$
        \ENDFOR
    \end{algorithmic}
\end{algorithm}

Fig.~\ref{fig:Clustering_and_Mapping} illustrates an example of the CNASA algorithm with $N_{geo}=2$ and $N_A / N_S = 2$.
As can be seen, the air nodes with diverse class distributions are assigned to the same satellites, with high communication efficiency achieved.
The pseudo-code of the algorithm is presented in Algorithm~\ref{alg:Clustering_and_mapping}. 
The time complexity of the algorithm is $O(N_A^2/N_S + N_S \cdot N_{geo}^2)$.
In practical deployment, the CNASA algorithm can be executed offline prior to the model training process by a ground station in a centralized manner. Following the virtual node strategy in~\cite{8932318}, the network topology including the logical satellite positions, the air nodes, and the IoRT devices is fixed, so the air node-logical satellite position assignment can be computed first, then we just need to fill the logical satellite positions with the nearest satellite in real-time to obtain 
the air node-satellite assignment. It is worth noting that the motion of satellites is not only highly dynamic but also periodic, so the network topology is also periodic. Therefore, the air nodes can send their geographic locations as well as the class distribution vectors to the ground station through the satellite network, and the air node-satellite assignments can be pre-calculated for a period and sent to all satellites over the satellite network by the ground station before the model training starts. For cases with the swift motion of satellites or a bigger inter-satellite model synchronization frequency (i.e., $\tau_2$), the connection link quality between the air nodes and their associated actual satellites may change greatly, we need to periodically apply the CNASA algorithm to improve the air node-satellite assignments.


In practical application, $N_{geo}$ is an adjustable hyperparameter to balance a trade-off between the time overhead and the class distribution.
In reality, the class distribution of devices that are geographically close may be similar\textcolor{blue}{~\cite{10.1007/978-3-030-58607-2_5}}, so setting $N_{geo}$ to a small value may not be conducive to the diversity of data.
On the contrary, if we assign a large value to $N_{geo}$, e.g., $N_S$, the data diversity can be satisfactory, but the geographic distance between two air nodes in the same cluster may become very far, leading to a large communication time cost.
Therefore, it is essential to decide the value of $N_{geo}$ carefully to derive a nice balance in practice.


%% file: Sections/Multi-orbit_Satellite_Network.tex
\section{Extension to Multi-Orbit Satellite Network}
{\label{sec:Multi-orbit}}

\begin{figure}[t!]
\centering
\includegraphics[width=0.43\textwidth]{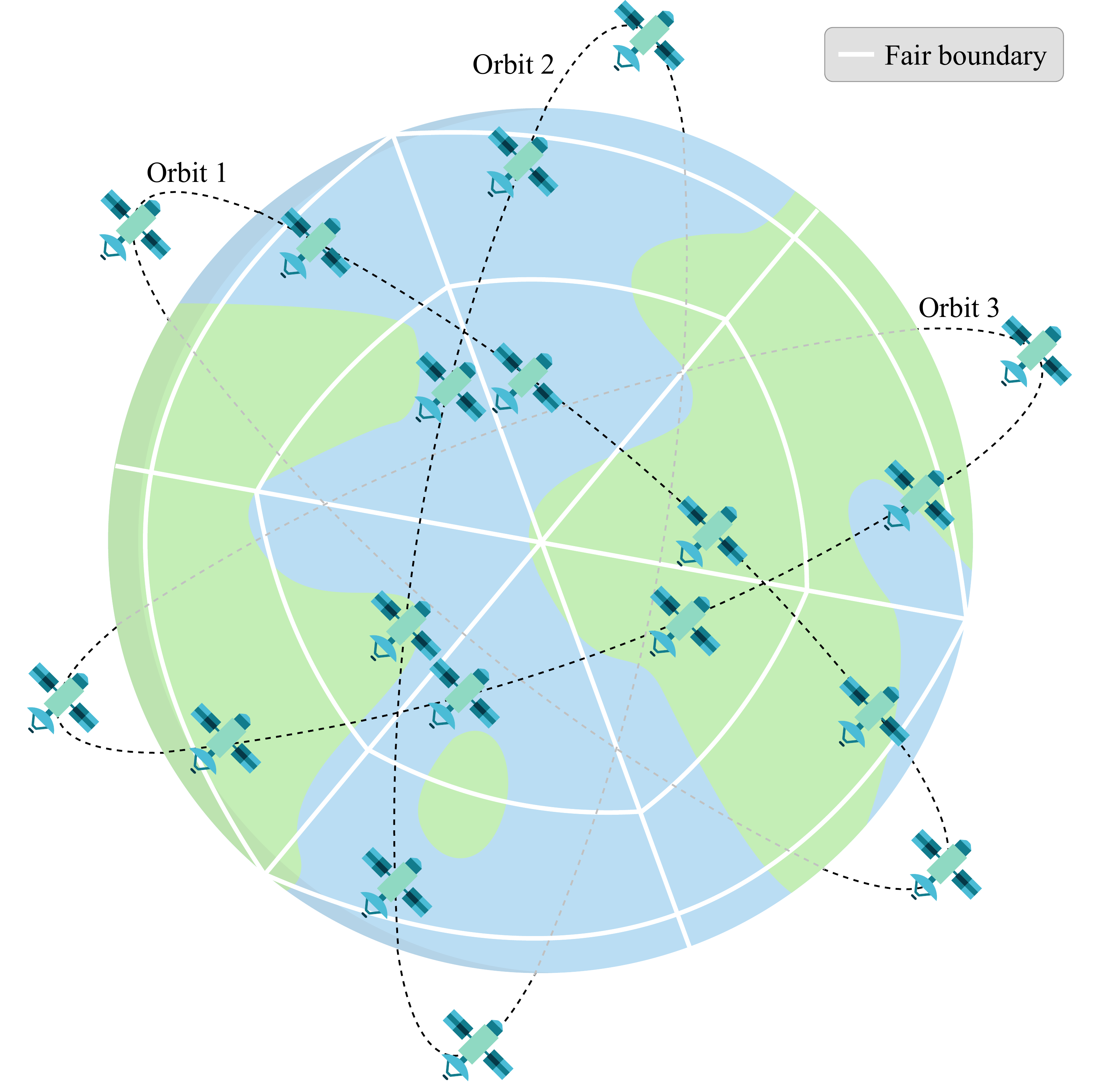}
\vspace{-5pt}
\caption{The fair boundary determined by Voronoi diagram.}
\label{fig:Voronoi}
\vspace{-10pt}
\end{figure}

Multi-orbit satellite network is a satellite network composed of multiple orbital planes at the same altitude, which can provide wider network coverage.
To make our OBL framework more generally applicable, we introduce an extension of OBL that can adapt to the multi-orbit satellite network.
However, the introduction of the multi-orbit satellite network confronts the challenges of more complex network topology, model synchronization among different orbits, and harder-to-measure model delivery time.
These problems mainly emerge in three parts, the determination of each satellite's coverage area, the inter-satellite model synchronization mechanism, and the geographical-distance-based division and cluster-to-satellite mapping in the CNASA algorithm.
Therefore, we elaborate our solution from these three parts.

\begin{figure*}[t!]
\vspace{-15pt}
    \centering
    \vspace{-15pt}
    \subfloat[Step 1: Intra-orbit Ring Allreduce.]{
    \vspace{-15pt}
    \label{fig:Multi_Allreduce_1}
    \includegraphics[width=0.43\textwidth]{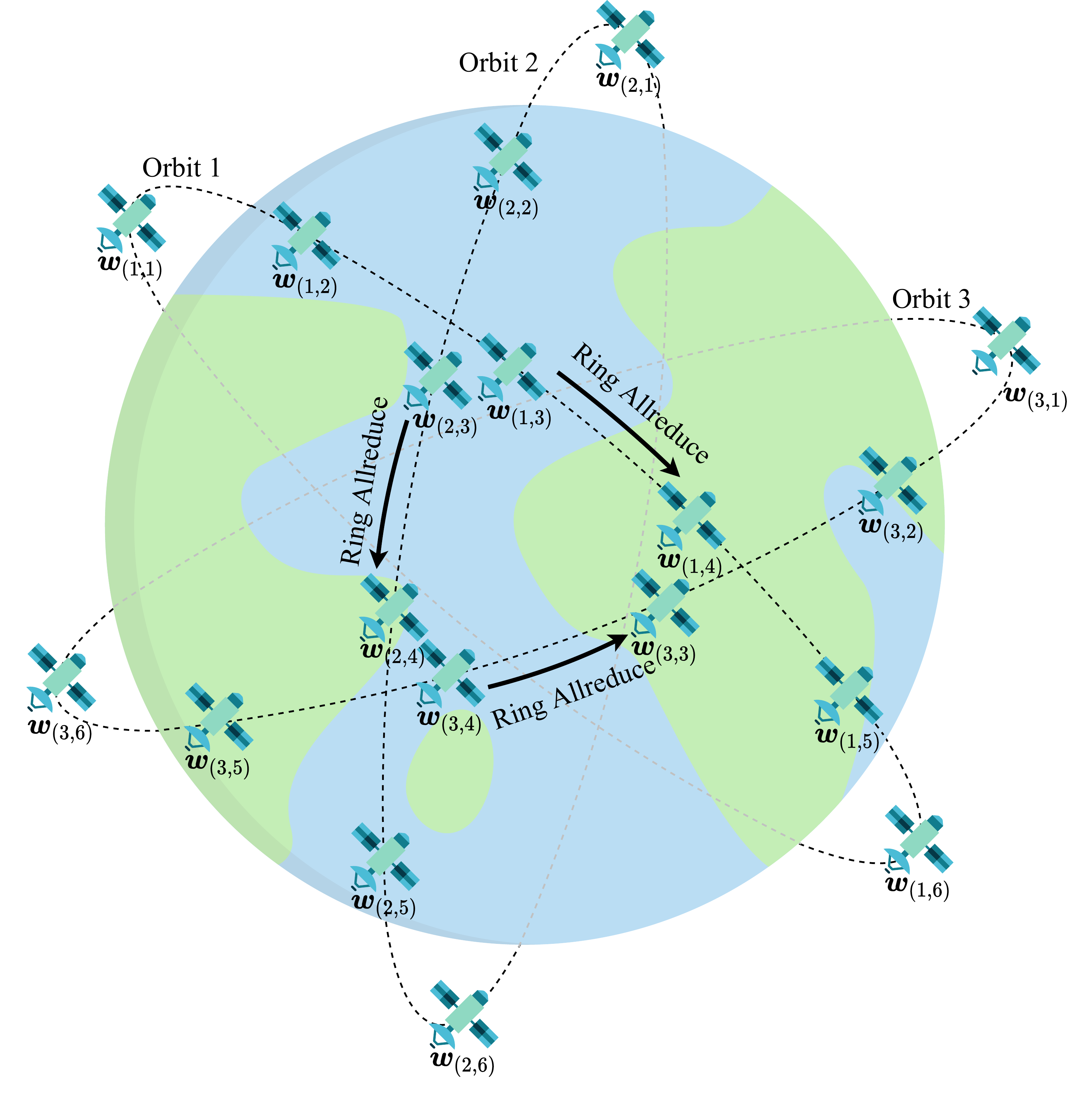}}
    \subfloat[Step 2: Inter-orbit Ring Allreduce.]{
    \vspace{-15pt}
    \label{fig:Multi_Allreduce_2}
    \includegraphics[width=0.43\textwidth]{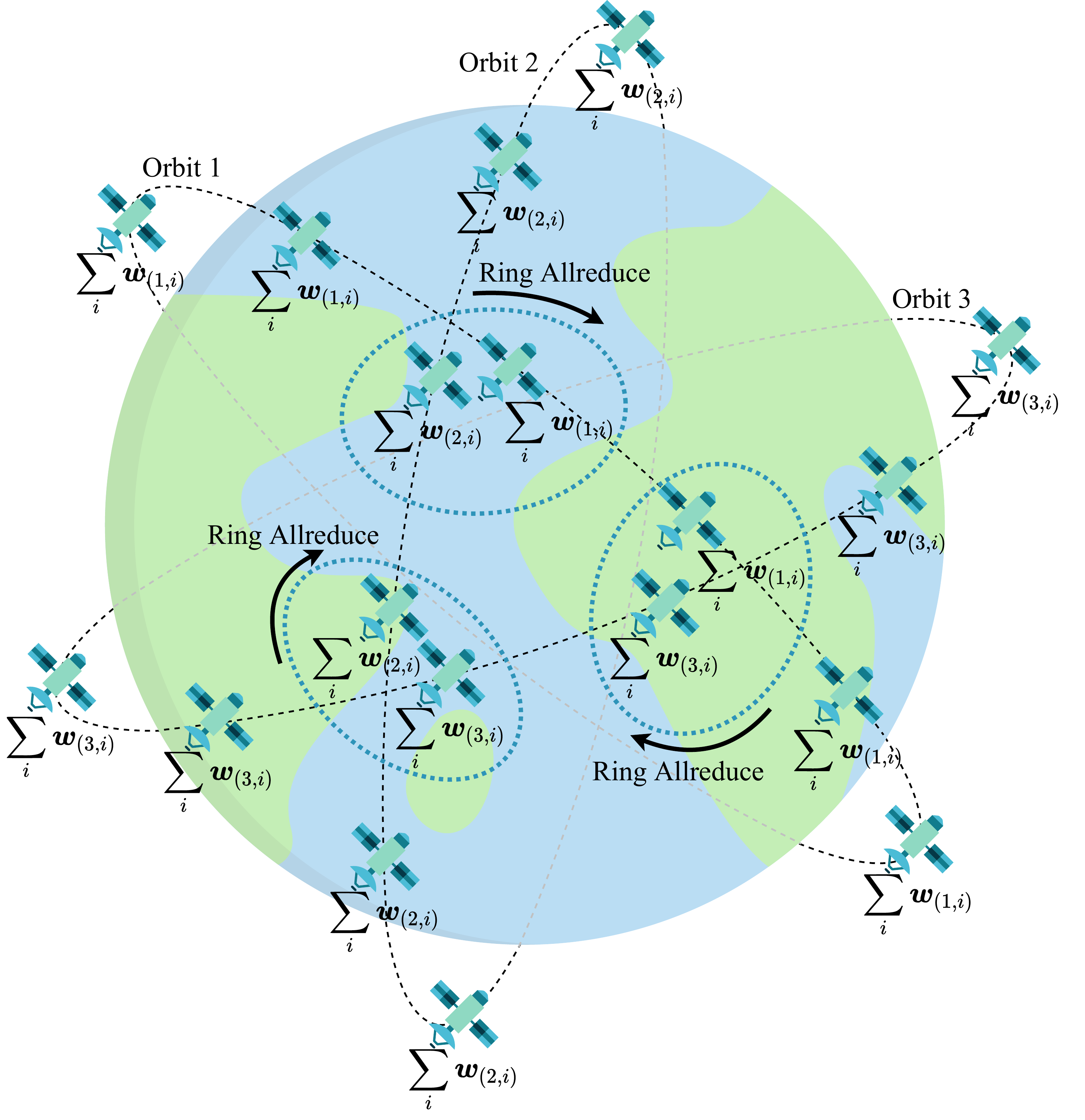}}
    \\
    \subfloat[Step 3: Intra-orbit Ring Allreduce.]{
    \label{fig:Multi_Allreduce_3}
    \includegraphics[width=0.43\textwidth]{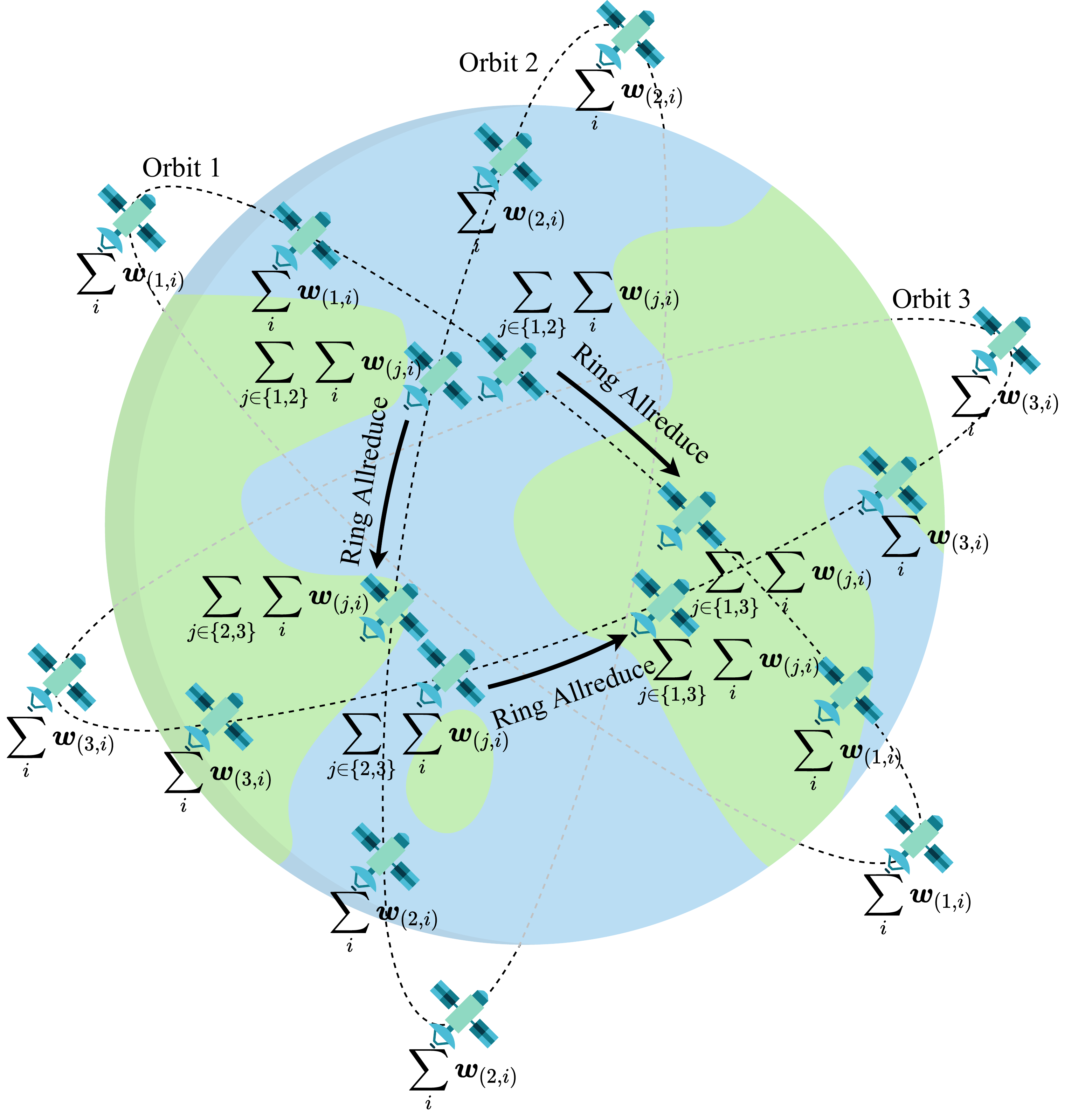}}
    \subfloat[The result.]{
    \label{fig:Multi_Allreduce_4}
    \includegraphics[width=0.43\textwidth]{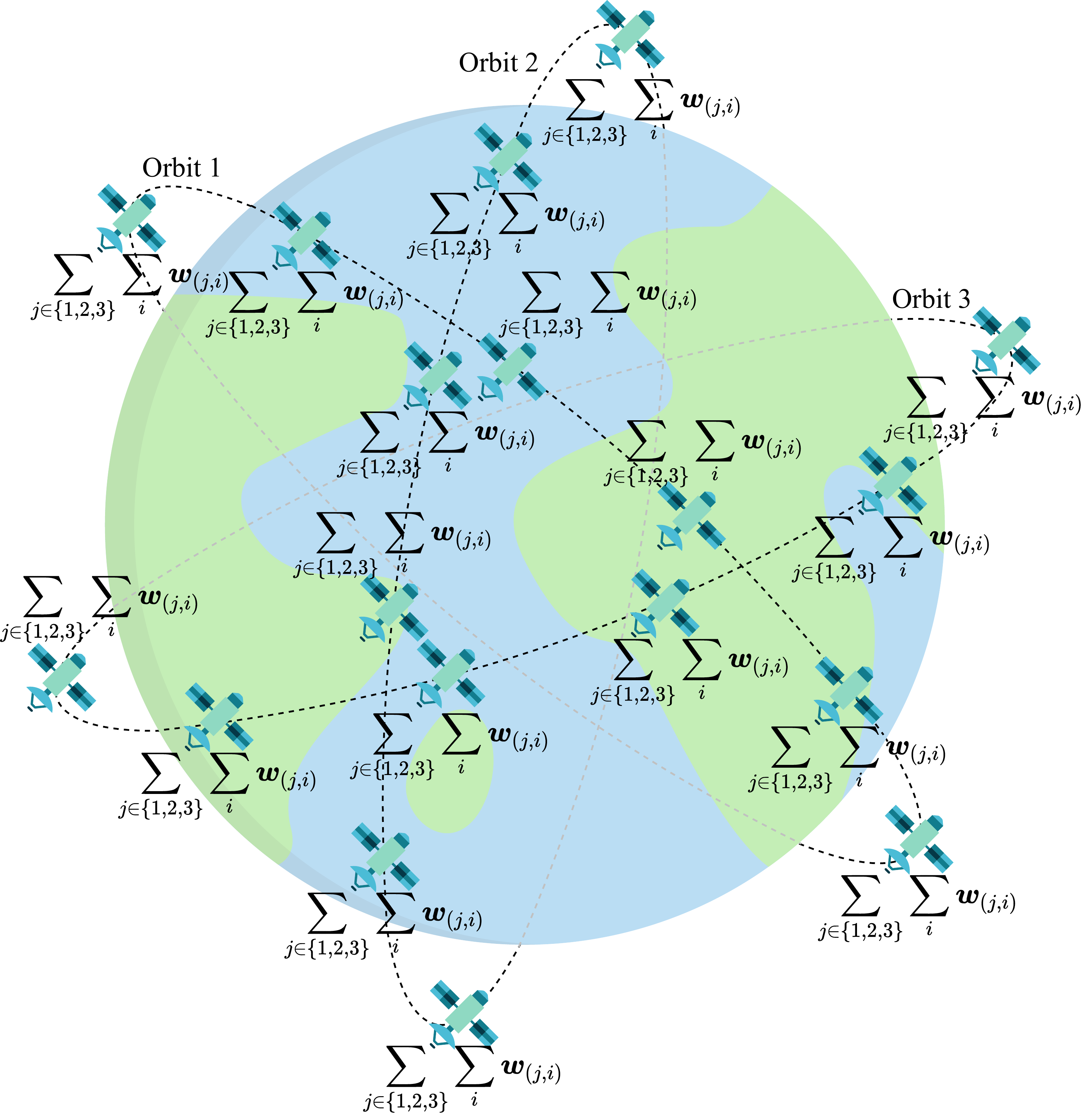}}
    \caption{The inter-satellite model synchronization process in multi-orbit satellite network.}
    \label{fig:Multi_Allreduce}
    \vspace{-15pt}
\end{figure*}

\subsection{Determination of Each Satellite's Coverage Area in Multi-Orbit Satellite Network}
{\label{subsec:Determination}}


Recall that in a single-orbit satellite network, we divide the coverage area of all satellites equally according to the geographical locations of all satellites to determine the access satellite of each air node.
However, in a multi-orbit satellite network, the coverage of satellites become more complicated as the air nodes may be covered by multiple satellites in different orbits.
To fairly identify the air nodes that each satellite can directly connect to, we first construct a Voronoi diagram~\cite{10.1007/978-3-642-13193-6_39} based on the position of all satellites.
Perpendicular-based division in the Voronoi diagram allows air nodes to be divided to the closest satellites, thus ensuring fairness.
Specifically, we calculate the positions of the satellites' projections on the earth's surface and draw the Voronoi diagram based on the positions, where each satellite corresponds to a region called a Voronoi cell, and the boundaries of all regions in the Voronoi diagram serve as the fair boundaries.
Fig.~\ref{fig:Voronoi} illustrates an example of the fair boundaries determined by the Voronoi diagram of satellites running on 3 orbits.
The fair boundaries marked by the white lines in Fig.~\ref{fig:Voronoi} can be used to divide the coverage area of all satellites to determine the air nodes that each satellite can directly connect to.

\begin{figure*}[t!]
\vspace{-0.5cm}
    \centering
    \subfloat[The inter-satellite link graph.]{
    \label{fig:Graph_Partition_1}
    \includegraphics[width=0.43\linewidth]{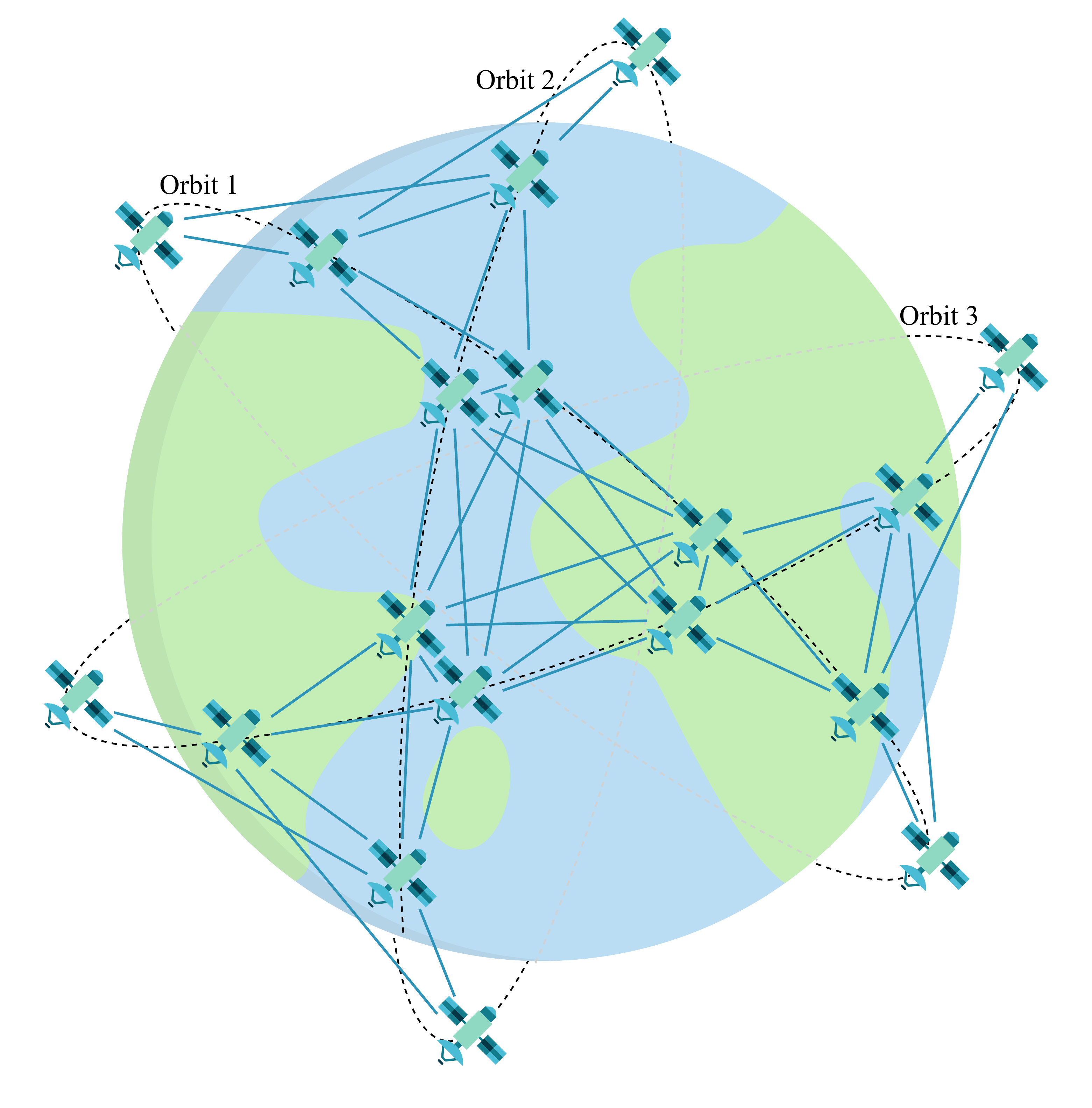}}
    \subfloat[The sub-graphs with $N_{geo}=3$.]{
    \label{fig:Graph_Partition_2}
    \includegraphics[width=0.43\linewidth]{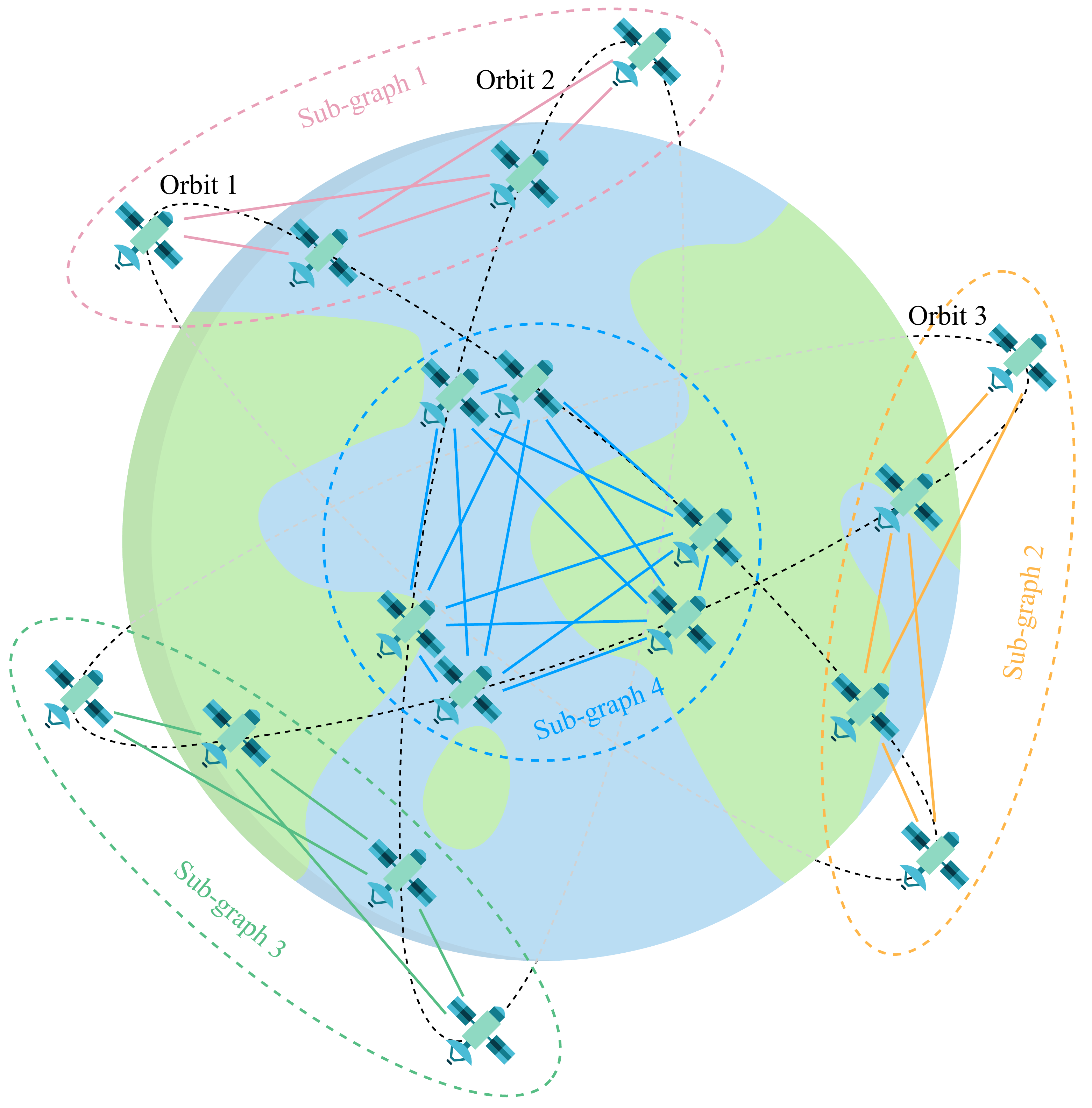}}
    \caption{The input and the output of the inter-satellite link graph partition algorithm.}
    \label{fig:Graph_Partition}
    \vspace{-10pt}
\end{figure*}

\subsection{Inter-Satellite Model Synchronization in Multi-Orbit Satellite Network}
In a multi-orbit satellite network,
the Ring Allreduce algorithm can still be used to synchronize the models within one orbit
due to the ring structure.
The key problem is, how to synchronize the models among different orbits after model synchronization within each orbit.
To address this issue, we first point out that at any time, there are inter-satellite links between any two orbits in the satellite network.
As an inherent characteristic, the satellite orbits are centered at the center of mass of the Earth and have equal radii, which ensures that there are intersections between any two satellite orbits, and the inter-satellite links between different orbits will exist around the intersections.
Although the satellites that establish inter-satellite links between different orbits at different times may change, as long as the inter-satellite links exist, model synchronization between orbits can be realized.
The Ring Allreduce algorithm can be used to efficiently synchronize the models of satellites around these intersections.
In fact, there is always more than one inter-satellite link between any two satellite orbits, but the model synchronization only needs to be performed once to avoid redundant synchronization.
In addition, if there are more than two satellites from different orbits around the intersection like the polar orbit satellite constellation, the Ring Allreduce algorithm can also be used to synchronize the models among all these satellites.
After model synchronization among different orbits, each orbit gets models from all the other orbits, so satellites within each orbit can perform the Ring Allreduce algorithm again to finish the inter-satellite model synchronization.
It is worth noting that the satellites not joining the model aggregation among different orbits just need to replace the chunks received when performing the Ring Allreduce algorithm.
For a more intuitive illustration, Fig.~\ref{fig:Multi_Allreduce} shows an example of the inter-satellite model synchronization running on 3 orbits, where $\boldsymbol w_{(j, i)}$ denotes the model of the $i$-th satellite in the $j$-th orbit before the inter-satellite model synchronization.

\subsection{The CNASA Algorithm in Multi-Orbit Satellite Network}
We extend the CNASA algorithm by modifying each step of it to accommodate the multi-orbit satellite network.
In a single orbit satellite network, the geographical-distance-based division of the CNASA algorithm ensures that the maximum number of model forwarding times associated with the relay satellites in one partition is smaller than $N_{geo}$, which is achieved by dividing the satellites into several isolated partitions.
However, in a multi-orbit satellite network, the topology graph of the inter-satellite links becomes more intricate, as shown in Fig.~\ref{fig:Graph_Partition_1}.
To limit the communication time cost of model uploading from air nodes to satellites, we propose an algorithm to divide the graph with all inter-satellite links into several sub-graphs to ensure that the maximum number of model forwarding times in each sub-graph, or the diameter of each sub-graph, is less than $N_{geo}$.
The algorithm is called the inter-satellite link graph partition and can be treated as an extension of the geographical-distance-based division for multi-orbit satellite network.
Specifically, the algorithm generates a sub-graph in each iteration, where the sub-graph is generated from a randomly picked satellite and expanded as the breadth-first search proceeds.
The distance (i.e., the number of hops) between any two satellites are computed at the beginning of each iteration to ensure that the diameter of the sub-graph is smaller than $N_{geo}$ when it is expanded.
The sub-graph is expanded until no more satellites can be added to the sub-graph, after which all satellites and their associated links are deleted from the original graph, and the next sub-graph are generated.
Fig.~\ref{fig:Graph_Partition_2} depicts an example of the partitioned sub-graphs with $N_{geo}$ equals 3, where the sub-graphs enclosed by dashed lines of different colors are the result of the inter-satellite link graph partition.
It can be seen that the diameter of each sub-graph is less than 3, consistent with the setting of the $N_{geo}$ value.
The pseudo-code of the algorithm is presented in Algorithm~\ref{alg:Graph_division}.
The time complexity of the algorithm is $O(N_S^4|E|)$, where $|E|$ denotes the number of inter-satellite links.
Similar to the CNASA algorithm, the inter-satellite link graph partition algorithm can also be executed offline by a powerful ground station in a centralized manner.

\begin{algorithm}[t]
\small
    \caption{Inter-Satellite Link Graph Partition}
    \label{alg:Graph_division}
    \begin{algorithmic}[1]
        \REQUIRE the inter-satellite link graph $\mathcal G$, the value that the diameter of each sub-graph must not exceed $N_{geo}$
        \ENSURE the set of sub-graphs $\mathcal S'=\{\mathcal S'_1, \mathcal S'_2, ...\}$
        \STATE Initialize an empty set of sub-graphs $\mathcal S'=\{\}$
        \WHILE {$\mathcal G$ is not empty}
            \STATE Compute the distance between any two satellites in $\mathcal G$
            \STATE Create an empty sub-graph $\mathcal S'_{new}$
            \STATE Randomly pick a satellite from $\mathcal G$ and add it to $\mathcal S'_{new}$
            \FOR {each satellite $S_{i}$ in $\mathcal S'_{new}$}
                \FOR {each adjacent satellite $S_{j}$ of $S_{i}$ in $\mathcal G$}
                    \IF {$S_{j}$ not in $\mathcal S'_{new}$ \AND the distance between $S_{j}$ and all satellites in $\mathcal S'_{new}$ is smaller than $N_{geo}$}
                        \STATE Add $S_{j}$ and the links between $S_{j}$ and its adjacent satellites in $\mathcal S'_{new}$ to $\mathcal S'_{new}$
                    \ENDIF
                \ENDFOR
            \ENDFOR
            \FOR {each satellite $S_{i}$ in $\mathcal S'_{new}$}
                \STATE Delete $S_{i}$ and its related links from $\mathcal G$
            \ENDFOR
            \STATE Add $\mathcal S'_{new}$ to $\mathcal S'$
        \ENDWHILE
    \end{algorithmic}
\end{algorithm}

After dividing the satellites into several partitions (i.e., sub-graphs), similar to Step~\ref{alg:Step_1} in Section~\ref{sec:Problem_Algorithm}, the air nodes can also be divided according to the coverage areas of satellites determined by the Voronoi diagram.
That is, an air node will be assigned to the partition that its access satellite belongs to according to the Voronoi diagram in Section~\ref{subsec:Determination} above.
Then, since the step of class-distribution-based clustering of the CNASA algorithm is not affected by the number of satellite orbits, this step does not need to be changed.
For the step of cluster-to-satellite mapping, the model delivery time from a source air node to a target satellite can be obtained based on the distance between the target satellite and the satellite that covers the source air node in the inter-satellite link graph, which has been computed in Algorithm~\ref{alg:Graph_division}.

%% file: Sections/Convergence_Analysis.tex
\section{Convergence Analysis of Olive Branch Learning}
\label{sec:Convergence_Analysis}

In this section, we study the convergence of OBL under general data distributions (e.g., non-IID distribution). For ease of exposition, we target on the convergence analysis of OBL (with single or multiple orbits) after the CNASA algorithm is applied such that the associations between the IoRT devices, air nodes, and satellites are fixed until the global FL model is obtained. We will consider the challenging scenario that during the training process, the associations between the IoRT devices, air nodes, and satellites can dynamically change in future work. Also, similar to many existing studies, we make the following assumption for the local loss function $ F_{k,i} $.

\begin{assumption}
\label{Assumption}
    For every $i$ and $k$, we assume:
    \begin{itemize}
        \item[$\triangleright$] (\textbf{Convexity}) $ F_{k,i} $ is convex.
        \item[$\triangleright$] (\textbf{Continuity}) $ F_{k,i} $ is $ \rho $-Lipschitz, i.e., for any $\boldsymbol w$ and $ \boldsymbol w' $, $ \lVert F_{k,i}(\boldsymbol w) - F_{k,i}(\boldsymbol w') \rVert \le \rho \lVert \boldsymbol w - \boldsymbol w' \rVert $.
        \item[$\triangleright$] (\textbf{Smoothness}) $ F_{k,i} $ is $ \beta $-smooth, i.e., for any $\boldsymbol w$ and $ \boldsymbol w' $, $ \lVert \nabla F_{k,i}(\boldsymbol w) - \nabla F_{k,i}(\boldsymbol w') \rVert \le \beta \lVert \boldsymbol w - \boldsymbol w' \rVert $.
    \end{itemize}
\end{assumption}



Remember that the calculations of the local model, the satellite model, and the global model are stated in Definition~\ref{def:Olive-Branch-Learning}.
We need to emphasize that the partial aggregation on the air nodes does not change the result of satellite model aggregation (because of the distributive law of multiplication), and the Ring Allreduce algorithm's convergence to a consistent weighted average model in the distributed setting has been proven in~\cite{allreduce}. Thus, we omit the description of the partial aggregation and the Ring Allreduce algorithm here. The gradient divergence of the local model and the satellite model is defined in Definition~\ref{def:GradientDivergence}, which reflects the impact of differences in data distribution across devices (i.e., IoRT devices or satellites) on Olive Branch Learning.

\begin{definition}[Gradient Divergence]
\label{def:GradientDivergence}
    For any $i$ and $k$, $ \delta_{k,i} $ is defined as the gradient difference between the $i$-th local loss and the $k$-th satellite loss, and $ \Delta_{k} $ is defined as the gradient difference between the $k$-th satellite loss and the global loss, which are expressed as
    \begin{equation}
    \small
    \begin{gathered}
    \label{eq:GradientDivergence}
        \delta_{k,i} = \max_{\boldsymbol w} \lVert \nabla F_{k,i}(\boldsymbol w) - \nabla F_{k}(\boldsymbol w) \rVert, \\
        \Delta_{k} = \max_{\boldsymbol w} \lVert \nabla F_{k}(\boldsymbol w) - \nabla F(\boldsymbol w) \rVert.
    \end{gathered}
    \end{equation}
    The overall local divergence $\delta$ and the overall satellite divergence $\Delta$ are formulated as
    \begin{equation}
    \small
    \begin{gathered}
    \label{eq:GradientDivergenceSum}    
        \delta = \sum_{ 1 \le k \le N_S }{ \sum_{ i \in \mathcal{N}_{k} }{ \frac{ |\mathcal{D}_{k,i}| }{ |\mathcal{D}| } \delta_{k,i} } }, ~~ 
        \Delta = \sum_{ 1 \le k \le N_S }{ \frac{ |\mathcal{D}_{k}| }{ |\mathcal{D}| } \Delta_{k} }.
    \end{gathered}
    \end{equation}
\end{definition}

The overall local divergence $\delta$ represents the overall gradient difference between all the IoRT devices' models and their respective satellites' models, and the overall satellite divergence $\Delta$ stands for the overall gradient difference between all the satellites' models and the global model.

To represent the divergence between a local model and a satellite model and between a satellite model and a global model, following the idea of~\cite{DBLP:journals/corr/abs-2012-03214}, we introduce the definition of virtual centralized learning in Definition~\ref{def:VirtualLearning}, where the training data for the satellite model and the global model is assumed to be stored centrally. We consider two types of intervals including a satellite interval $ [s] \triangleq [ (s-1)\tau_1, s\tau_1 ] $ that denotes a time interval between two consecutive satellite aggregations, and a global interval $ [g] \triangleq [ (g-1)\tau_1\tau_2, g\tau_1\tau_2 ] $ that denotes a time interval between two consecutive inter-satellite model synchronizations.


\begin{definition}[Virtual Centralized Learning]
\label{def:VirtualLearning}
    For any $t$, $k$, $[s]$, and $[g]$, the virtual satellite model $ \boldsymbol v^{(t)}_{k} $ and the virtual global model $ \boldsymbol v^{(t)} $ are synchronized with the satellite model $ \boldsymbol w^{(t)}_{k} $ and the global model $\boldsymbol w^{(t)}$ at the beginning of each interval, respectively, and trained on the centralized data examples of $\mathcal{N}_{k}$ and $\mathcal{N}$, respectively, which is represented as
    \begin{equation}
    \small
    \label{eq:v_central}
    \begin{aligned}
        \boldsymbol v^{(t)}_{k} & = \begin{dcases}
            \boldsymbol w^{(t)}_{k} & \text{if} ~ t = (s-1)\tau_1, \\
            \boldsymbol v^{(t-1)}_{k} - \eta \nabla F_{k}(\boldsymbol v^{(t-1)}_{k}) & \text{if} ~ t \neq (s-1)\tau_1,
        \end{dcases} \\
        \boldsymbol v^{(t)} & = \begin{dcases}
            \boldsymbol w^{(t)} & \text{if} ~ t = (g-1)\tau_1\tau_2, \\
            \boldsymbol v^{(t-1)} - \eta \nabla F(\boldsymbol v^{(t-1)}) & \text{if} ~ t \neq (g-1)\tau_1\tau_2.
        \end{dcases}
    \end{aligned}
    \end{equation}
\end{definition}

Based on the assumption and definitions above, for Olive Branch Learning, we derive the convergence bound between the global model and the virtual global model by Theorem~\ref{def:Conv-Olive-Branch-Learning}. The detailed proof is similar to that in~\cite{DBLP:journals/corr/abs-2012-03214} and hence is omitted here due to space limit.  

\begin{theorem}
\label{def:Conv-Olive-Branch-Learning}
The convergence bound of Olive Branch Learning is given by
\begin{equation}
\small
\label{eq:Conv-Olive-Branch-Learning}
    \lVert F(\boldsymbol w^{(t)}) - F(\boldsymbol v^{(t)}) \rVert \le \frac{ \rho }{ \beta }( \delta h(\tau_1) + \Delta h(\tau_1\tau_2) ),
\end{equation}
where $h(t) = ( \eta\beta + 1 )^{t} - 1$.
\end{theorem}


From Theorem~\ref{def:Conv-Olive-Branch-Learning}, we conclude that the convergence bound of OBL is proportional to $\delta$ and $\Delta$, where the coefficient of $\Delta$ is bigger than that of $\delta$, since $h(\tau_1\tau_2) > h(\tau_1)$, in other words, it's more efficient to decrease $\Delta$ to get a smaller convergence bound of OBL. The CNASA algorithm in our OBL framework tends to cluster the air nodes so that the average class distribution of air nodes within each cluster is close to the global class distribution, thus decreasing the overall satellite divergence $\Delta$ and facilitating the convergence of the global model.

%% file: Sections/Performance_Evaluation.tex

\begin{figure*}[t!]
    \centering
    \begin{minipage}[t]{0.485\textwidth}
        \centering
        \includegraphics[width=1.0\textwidth]{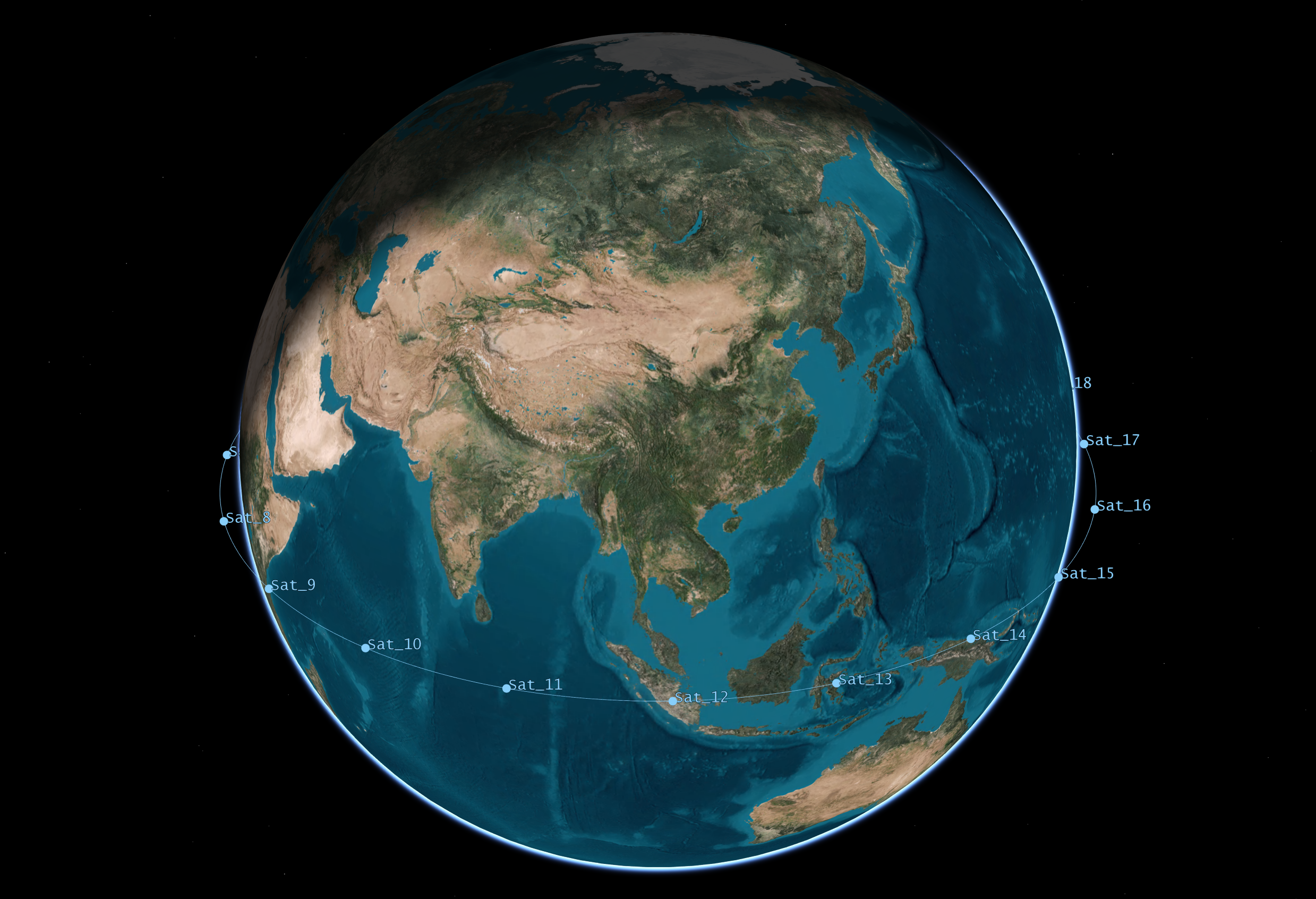}
        \caption{Schematic diagram of the single-orbit satellite network.}
        \vspace{-15pt}
        \label{fig:Single_orbit}
    \end{minipage}
    \hspace{0.05in}
    \centering
    \begin{minipage}[t]{0.485\textwidth}
        \centering
        \includegraphics[width=1.0\textwidth]{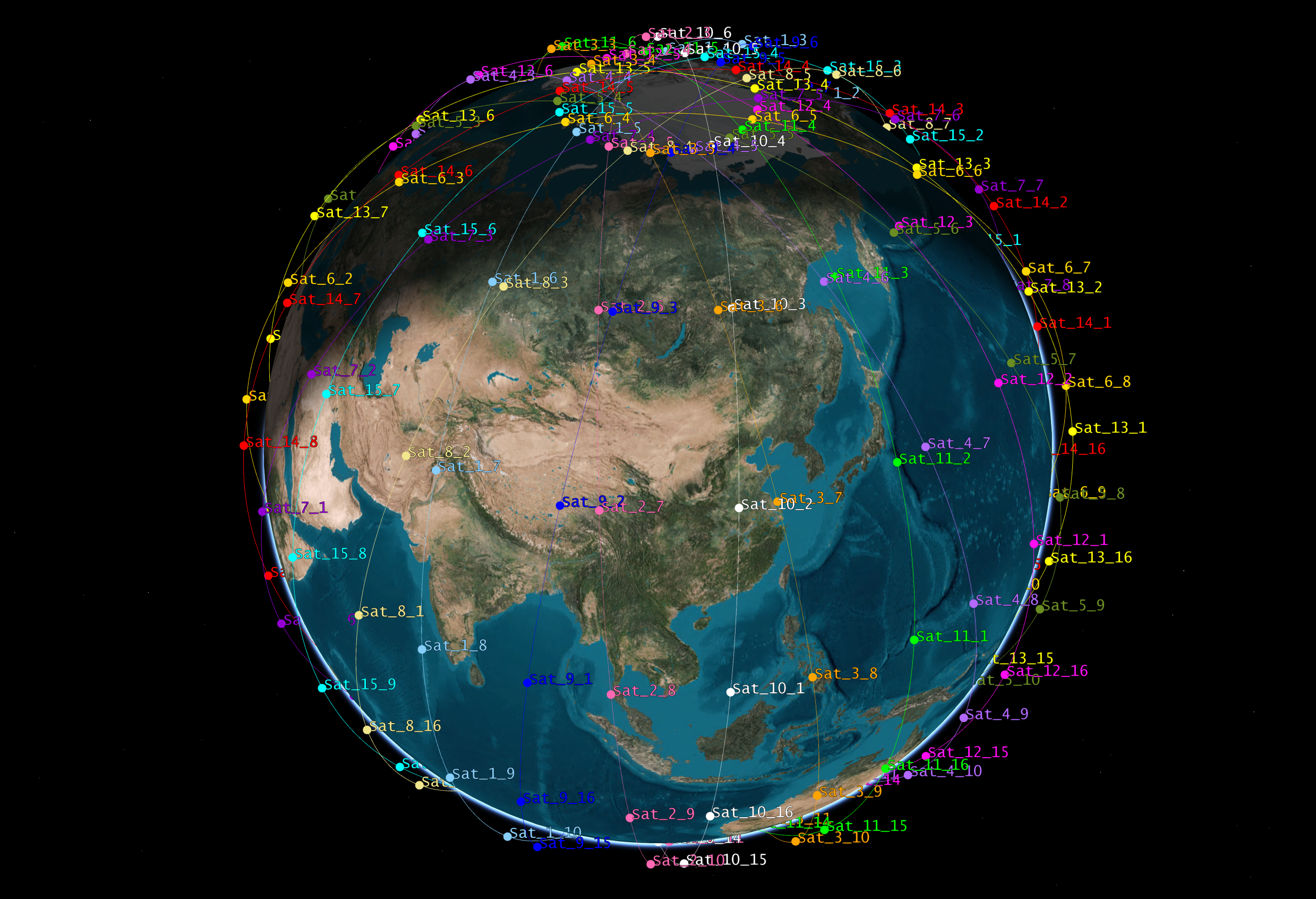}
        \caption{Schematic diagram of the multi-orbit satellite network.}
        \vspace{-15pt}
        \label{fig:Multi_orbit}
    \end{minipage}
\end{figure*}

\section{Performance Evaluation}
{\label{sec:Performance_Evaluation}}



\subsection{Experiment Setup} 

\subsubsection{Network settings} We consider a space-air-ground integrated network with 20 low earth orbit (LEO) satellites, 100 air nodes, and 200 IoRT devices, where the satellite network is illustrated in Fig.~\ref{fig:Single_orbit}.
The LEO satellites are distributed in an equatorial orbital plane at an altitude of about 330 km, the air nodes are evenly distributed at a height of 100 m above the equator, and each air node is responsible for the model aggregation of two IoRT devices without overlapping.
The connectivity data between satellites and air nodes is obtained from Satellite Tool Kit (STK) simulation software.
Referring to~\cite{9372909, electronics8111247}, we set the propagation delay and the data rate of the links, and the computing capabilities of devices, which are listed
in Table~\ref{table:network_settings}.

\setlength\tabcolsep{4 pt}
\begin{table}[t!]
    \centering
    \caption{Network settings.}
    \label{table:network_settings}
    \begin{tabular}{p{6.3cm}c}
        \hline
        \textbf{Parameter} & \textbf{Value}\\
        \hline
        Number of LEO satellites & 20 \\
        Propagation delay between an IoRT device and its access air node & 5 ms \\
        Number of air nodes & 100 \\
        Propagation delay between an air node and its access satellite & 5 ms \\
        Number of IoRT devices & 200 \\
        Total bandwidth of an satellite available for its associated air nodes & 6000 Mbps\\
        Altitude of orbital plane & 330 km \\
        Total bandwidth of an air node available for its accessed IoRT devices & 32 Gbps \\
        Flying height of air nodes & 100 m \\
        Bandwidth of the data link between two neighboring satellites & 30 Gbps \\
        Propagation delay between two neighboring satellites & 20 ms \\
        Computing capabilities of the IoRT devices, the air nodes, and the satellites & 0.665 TFLOPS \\
        \hline
    \end{tabular}
    \vspace{-10pt}
\end{table}

\subsubsection{Training settings} For the federated learning task, we focus on the image classification task and use the popular model training datasets MNIST, Fashion-MNIST, and CIFAR-10, which are widely used for testing in many federated learning works~\cite{pmlr-v54-mcmahan17a, 9148862, 9337204}, to evaluate the performance of the CNASA algorithm.
Since our OBL framework does not impose any restrictions on training models for practical applications, it can meet the actual demands of SAGIN. Due to the page limit and the lack of the datasets of space applications, we will consider evaluating more relevant training tasks in future work. Each dataset consists of a training set for model training and a test set for model evaluation, and the specific explanation are as follows:

\begin{itemize}
\item[$\triangleright$] \textbf{MNIST}~\cite{mnist} is a classic hand-written digit classification dataset with 10 classes. The learning model used for MNIST is a convolution neural network (CNN) with 2 convolutional layers and 2 fully connected layers, with 21,840 trainable parameters.
\item[$\triangleright$] \textbf{Fashion-MNIST}~\cite{fashion-mnist} is a dataset of article images with 10 classes. The model used for Fashion-MNIST is a CNN with 2 convolutional layers and 2 fully connected layers, with 421,642 trainable parameters.
\item[$\triangleright$] \textbf{CIFAR-10}~\cite{cifar-10} is an object recognition dataset with 10 classes. The model used for CIFAR-10 is a CNN with 9 convolutional layers, with one Group Normalization~\cite{Wu_2018_ECCV} layer between every two convolutional layers. The number of trainable parameters is 1,369,738.
\end{itemize}

\begin{figure*}[t!]
    \centering
    \includegraphics[width=0.32\textwidth]{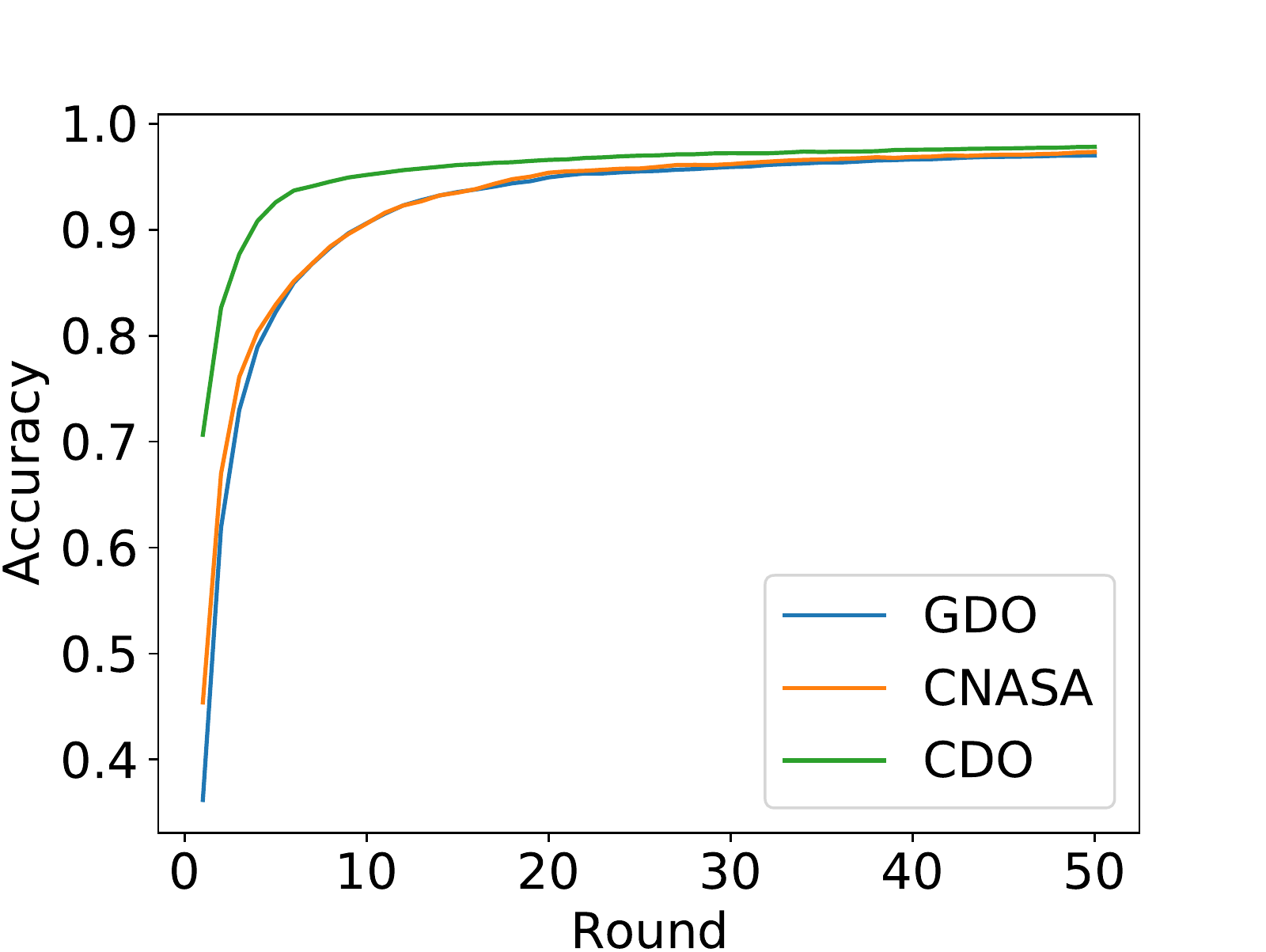}
    \includegraphics[width=0.32\textwidth]{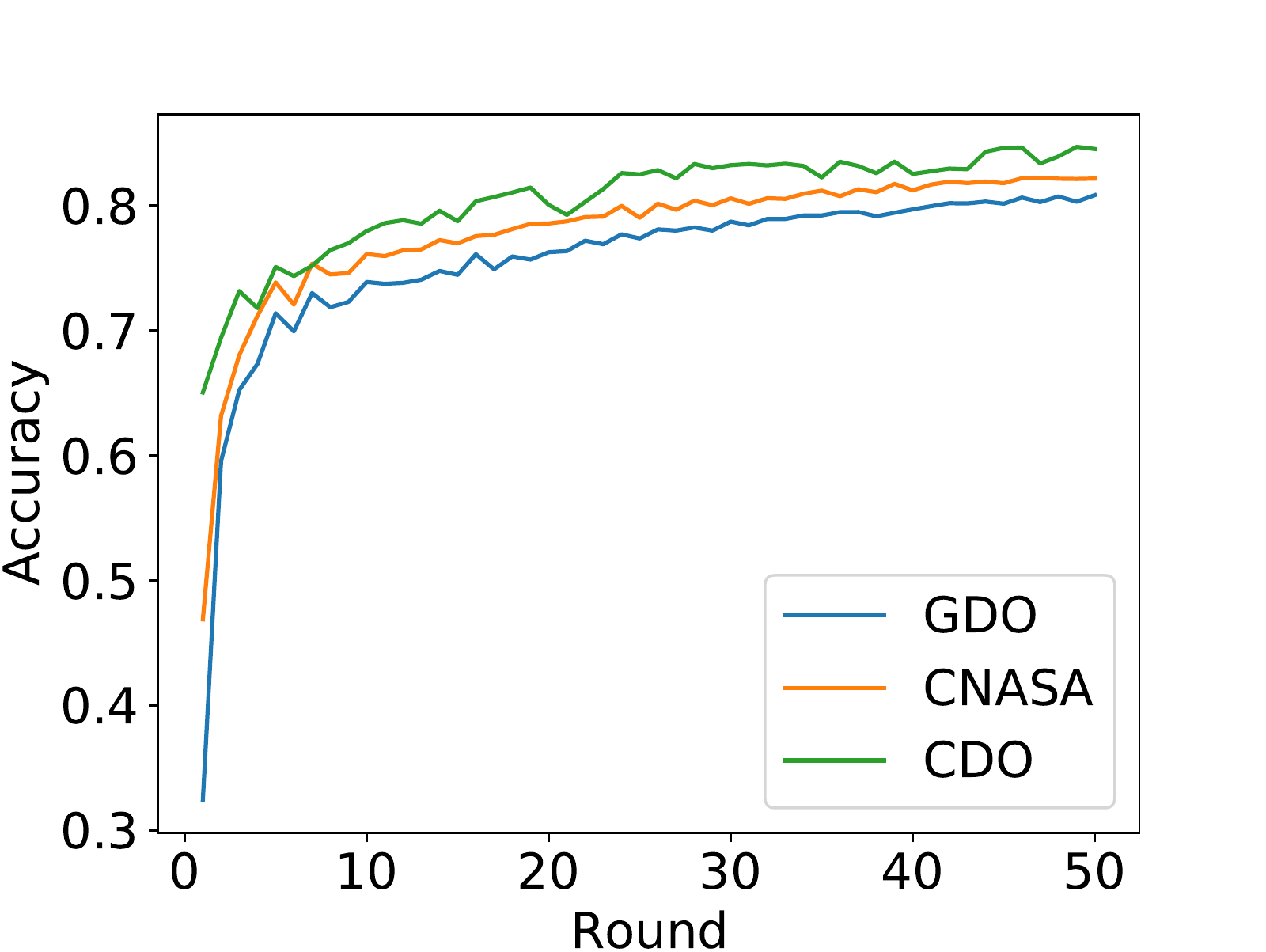}
    \includegraphics[width=0.32\textwidth]{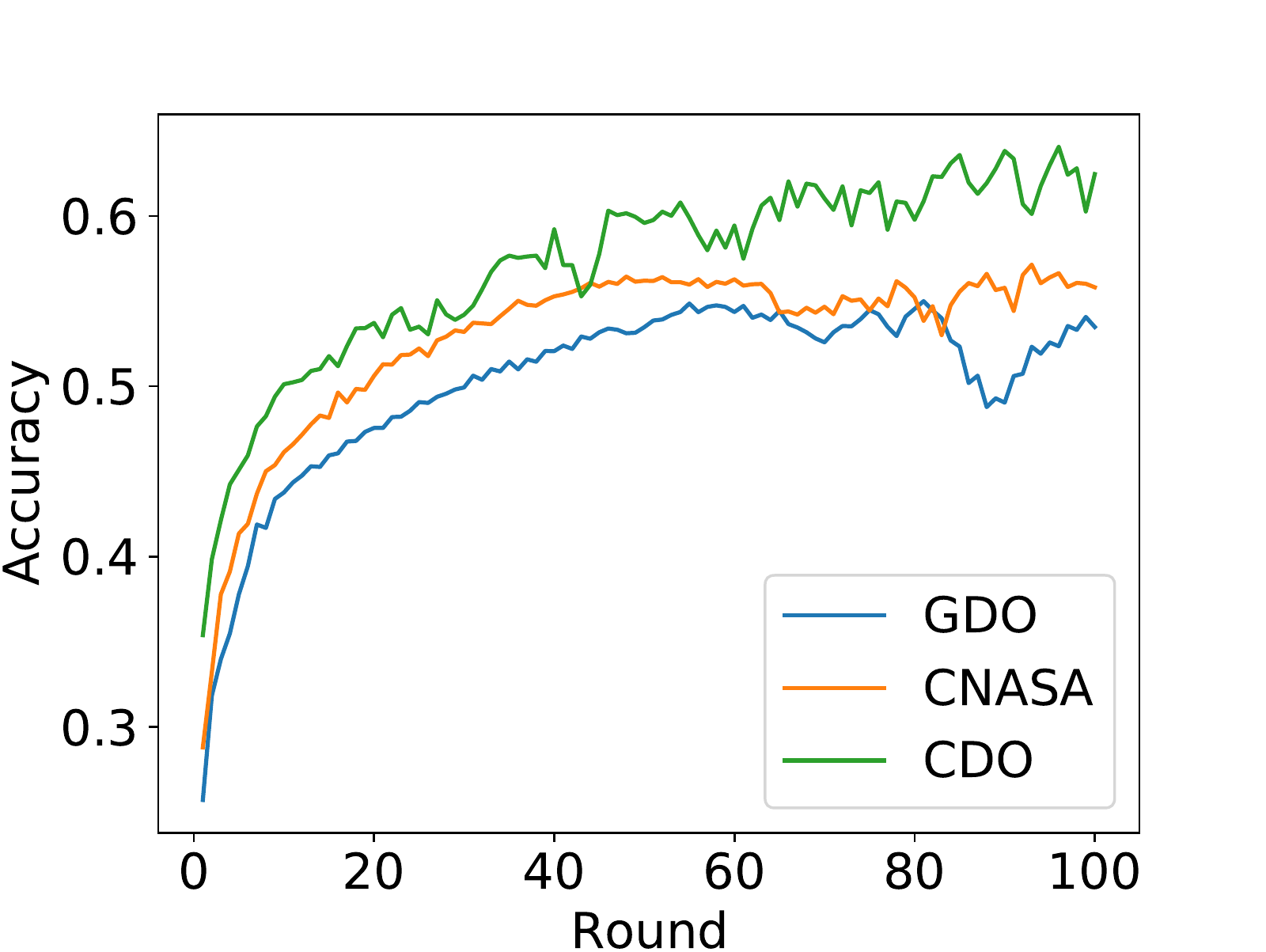}
    \vspace{-5pt}
    \caption{Test accuracy varies with global training rounds on MNIST (left), Fashion-MNIST (center), and CIFAR-10 (right).}
    \label{fig:acc_round}
    \vspace{-5pt}
\end{figure*}

\begin{figure*}[t!]
    \centering
    \includegraphics[width=0.32\textwidth]{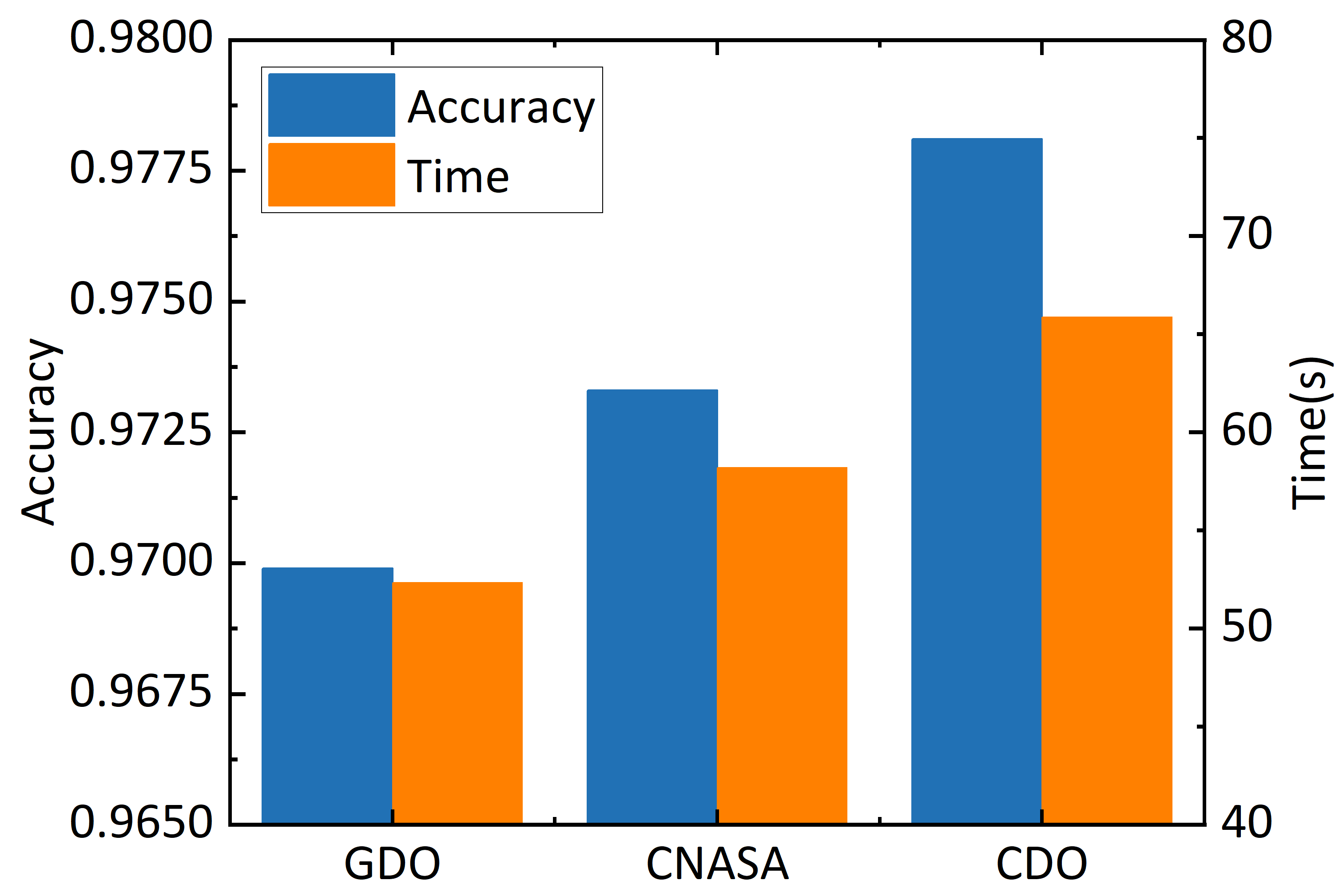}
    \includegraphics[width=0.31\textwidth]{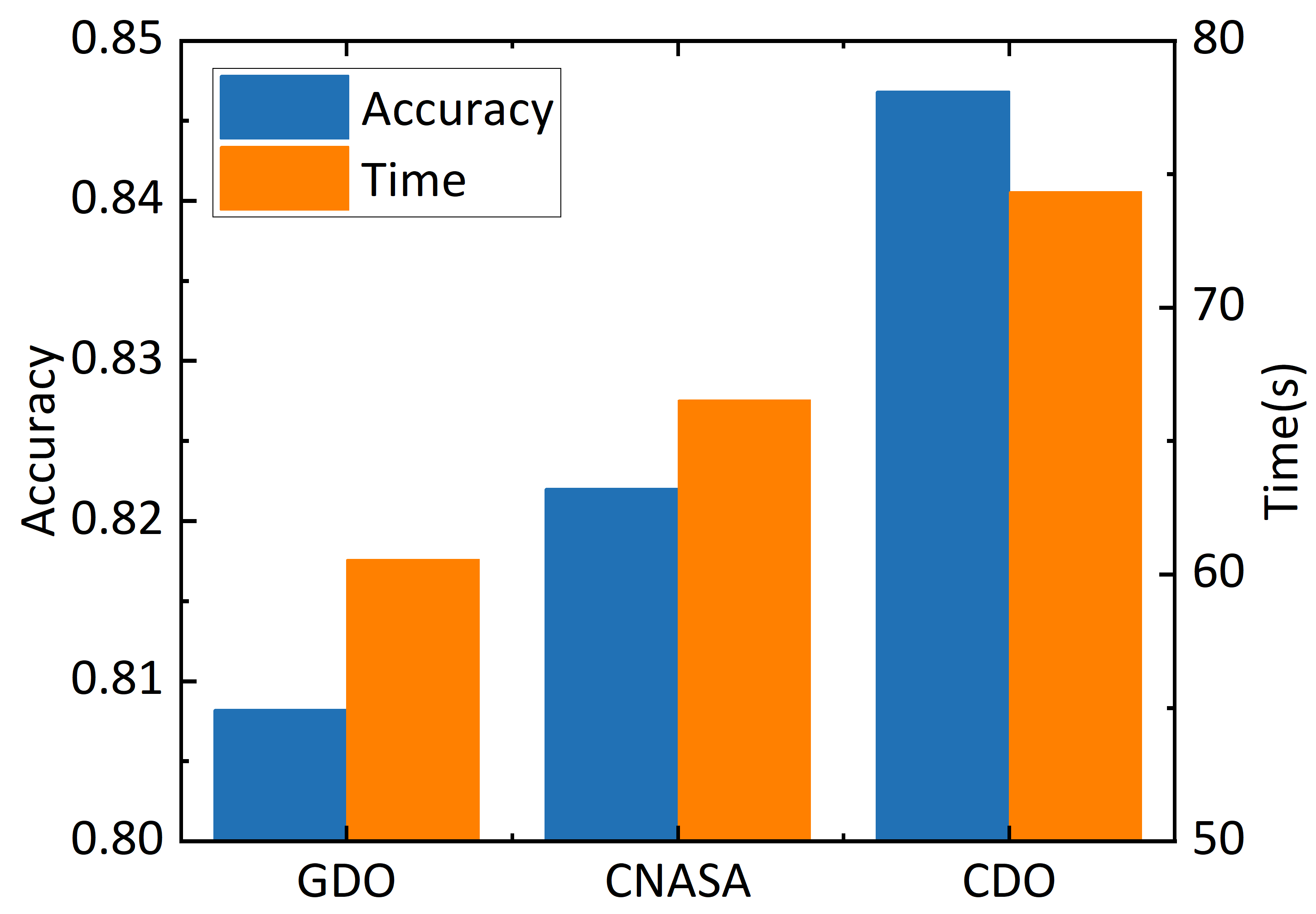}
    \includegraphics[width=0.315\textwidth]{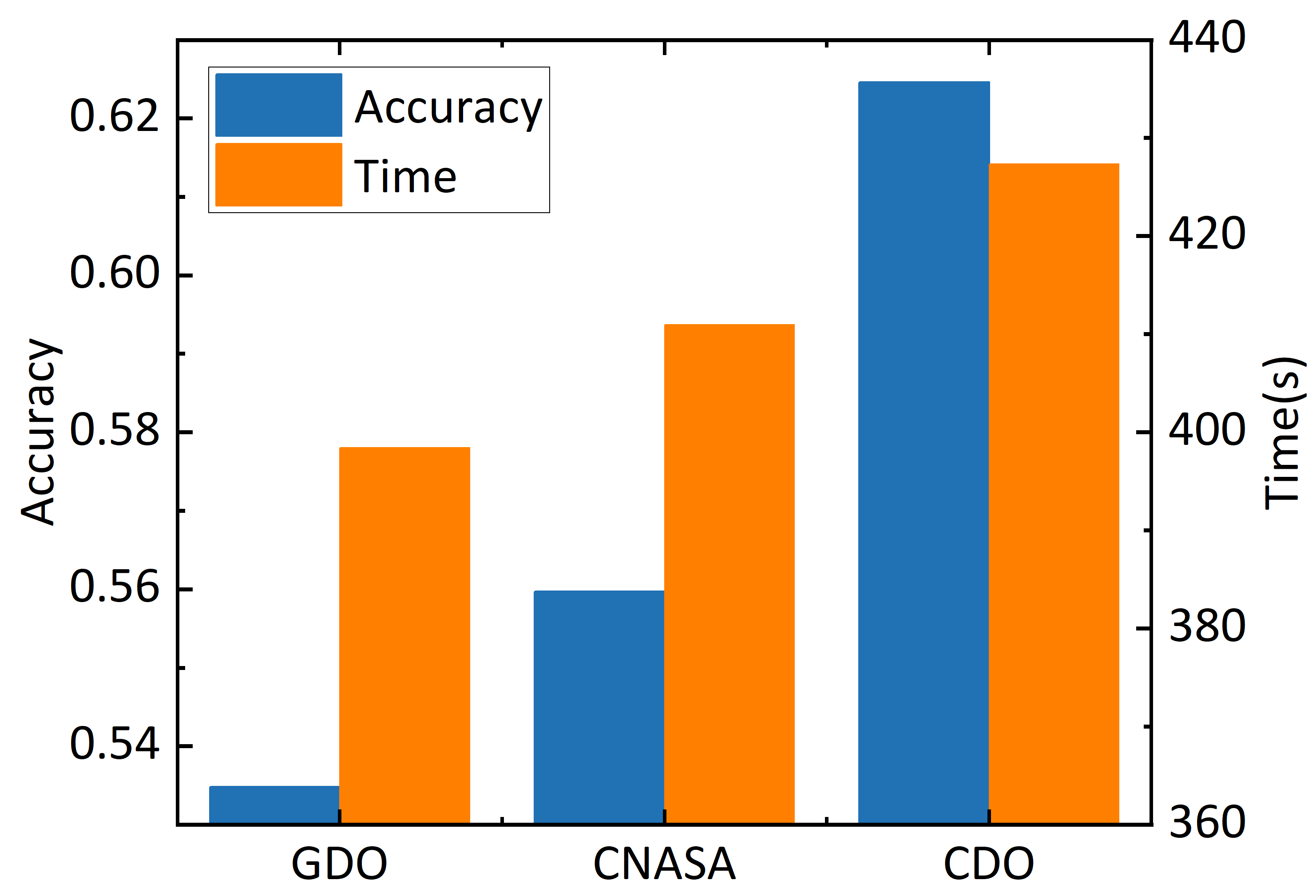}
    \vspace{-5pt}
    \caption{Test accuracy and time cost after 50, 50, and 100 global training rounds on MNIST (left), Fashion-MNIST (center), and CIFAR-10 (right), respectively.}
    \label{fig:acc_time}
    \vspace{-10pt}
\end{figure*}

Each IoRT device possesses its own dataset with the same size but different class distribution. To mimic the challenging non-IID scenario, we assign each IoRT device samples of 2 classes,
which will bring a significant negative impact on the global model accuracy.
The class distribution of geographically close devices is similar. 

\subsubsection{Comparison settings}
Our proposed OBL framework is the first attempt to deploy FL into SAGIN scenarios by leveraging the unique characteristics of SAGIN such as system hierarchy and the specific topologies of different layers. Current popular FL algorithms (e.g., FedAT~\cite{10.1145/3458817.3476211} and Astraea~\cite{9141436}) can not be directly applied to the SAGIN scenarios due to the complex topology structure of SAGIN. Hence, we design OBL+GDO and OBL+CDO methods, deriving from FedAT and Astraea, respectively, to adapt to SAGIN scenarios, for comparison with our proposed OBL+CNASA algorithm. Our evaluation metrics include overall time cost of model training and final global model accuracy. The comparison methods are listed as follows:

\begin{itemize}
\item[$\triangleright$] Geographic Distance Only (GDO): The air nodes only send the models to their access satellites, which does not rely on relay satellites to help forward the models.
\item[$\triangleright$] Class Distribution Only (CDO): The air nodes with large differences in class distribution vectors across the globe send their models to the same satellite, which may require a huge amount of relay satellites to help forward the models.
\end{itemize}


\subsection{Experiment Results}

\subsubsection{Evaluation on single-orbit satellite network} 
For reference, we first set $97\%$, $82\%$ and $55\%$ as the acceptable test accuracy for MNIST, Fashion-MNIST and CIFAR-10 datasets, respectively.
As illustrated in Fig.~\ref{fig:acc_round}, the models trained by different methods (e.g., GDO, CNASA and CDO) on the MNIST and Fashion-MNIST datasets converge fast in the early stages of learning and stabilize to a stationary point after dozens of rounds. As a more challenging classification dataset, CIFAR-10 consumes more global training rounds to achieve a relatively stable model accuracy, and the final model accuracy is significantly lower than that on the other two datasets. Among these methods, the CDO method achieves the highest test accuracy throughout the learning process on all datasets, which is because, in the CDO method, each satellite can perform model aggregation for air nodes with the most diverse class distributions around the world, neglecting the air nodes' geographic distance. However, as shown in Fig.~\ref{fig:acc_time}, CDO method requires more training time for model convergence, while our proposed CNASA algorithm can strike a nice balance between model accuracy and training efficiency. For example, comparing with CDO method, CNASA algorithm can reduce more than $14\%$ training time cost with only $0.5\%$ model accuracy degradation in the MNIST dataset.

\subsubsection{Effect of \texorpdfstring{$N_{geo}$}{N\_geo} value}
\begin{figure*}[t!]
    \flushright
    \includegraphics[width=0.32\textwidth]{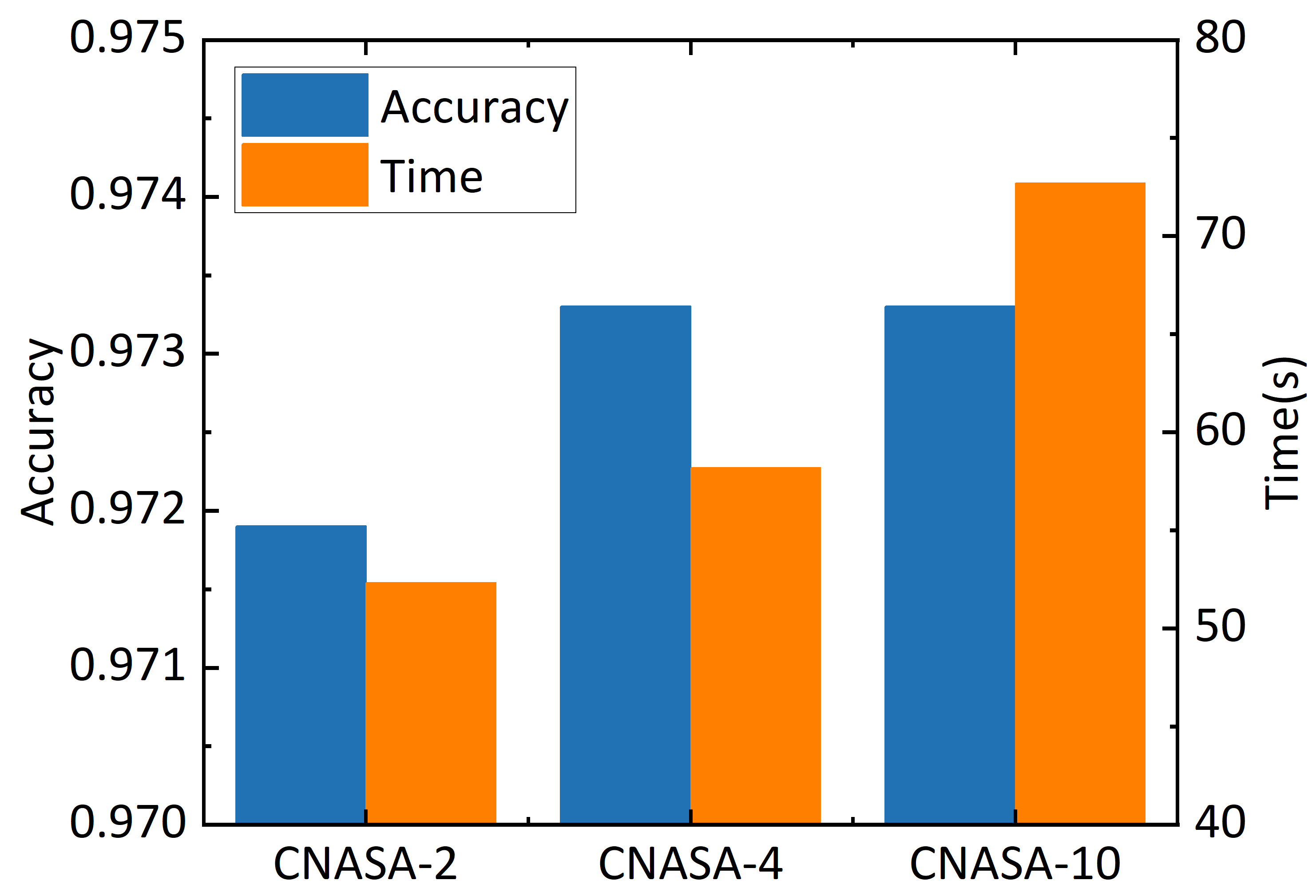}
    \includegraphics[width=0.32\textwidth]{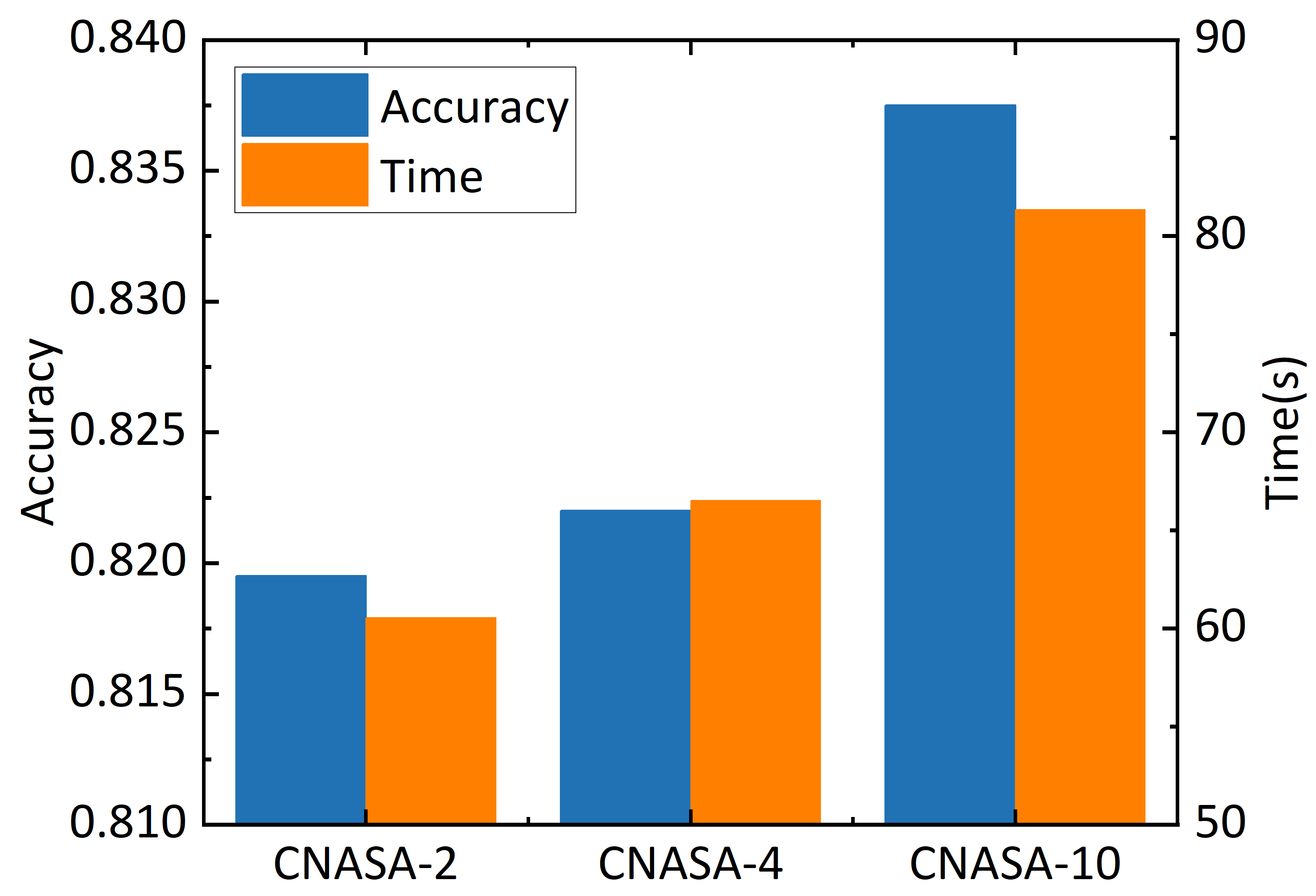}
    \includegraphics[width=0.32\textwidth]{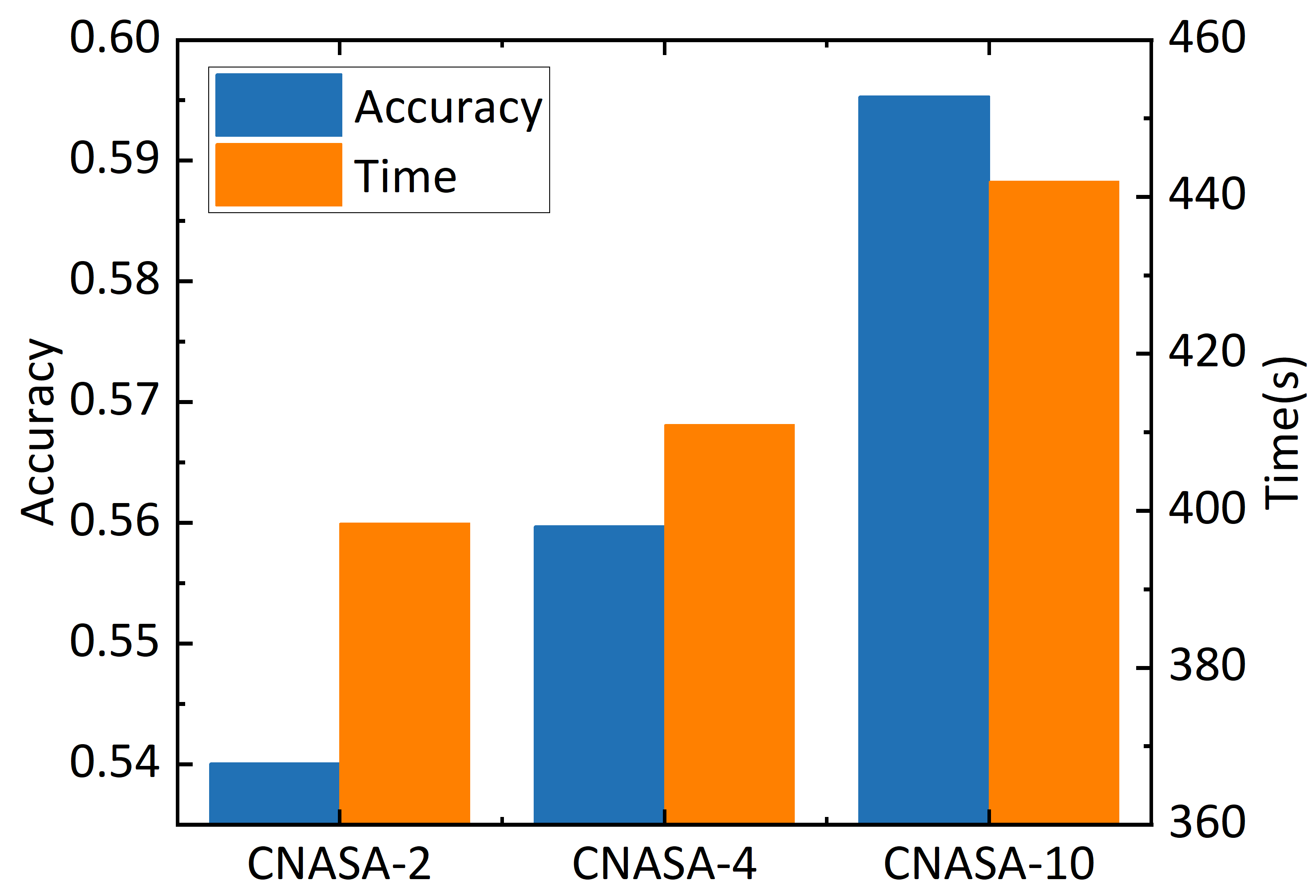}
    \vspace{-5pt}
    \caption{Effect of different $N_{geo}$ values on on MNIST (left), Fashion-MNIST (center), and CIFAR-10 (right).}
    \label{fig:N_geo}
    \vspace{-5pt}
\end{figure*}

We use CNASA-2, CNASA-4, and CNASA-10 to denote the CNASA algorithm with $N_{geo}=2$, $N_{geo}=4$, and $N_{geo}=10$, respectively. The results are shown in Fig.~\ref{fig:N_geo}.
For each dataset, as the value of $N_{geo}$ in the CNASA algorithm increases, the test set accuracy of the model increases, while the time cost also increases.
This is because the larger the value of $N_{geo}$, the larger the data diversity in each cluster, which is more conducive to improving the accuracy of the model.
However, a larger value of $N_{geo}$ may also incur a large number of model forwarding times, thus increasing the time cost.
As a result, $N_{geo}$ acts as a hyperparameter that can be carefully tuned to strike a nice balance between global model accuracy and time cost and plays the role of $\gamma$ in \eqref{eq:objective_function}.
That is, if we prefer a model with higher accuracy, we can assign a relatively high value to $N_{geo}$, 
and vice versa.

\subsubsection{Effect of inter-satellite model synchronization frequency}
As shown in Fig.~\ref{fig:tau_2}, both the global model accuracy and the time cost increase as the value of $\tau_2$ enlarges, which is because a larger $\tau_2$ value allows more satellite model aggregations within one global training round, thus favoring more fully trained models, but also increasing the duration of each global training round.

\subsubsection{Effect of non-IID degree}
We fix $N_{geo} = 4$ in our algorithm, and evenly distribute the CIFAR-10 dataset among 200 IoRT devices in four different ways, i.e., Non-IID-1, Non-IID-2, Non-IID-5 and IID, representing that each IoRT device possesses data with one, two, five and ten classes, respectively. 
%
\begin{figure*}[t!]
    \flushleft
    \begin{minipage}[t]{0.321\textwidth}
        \centering
        \includegraphics[width=1.0\textwidth]{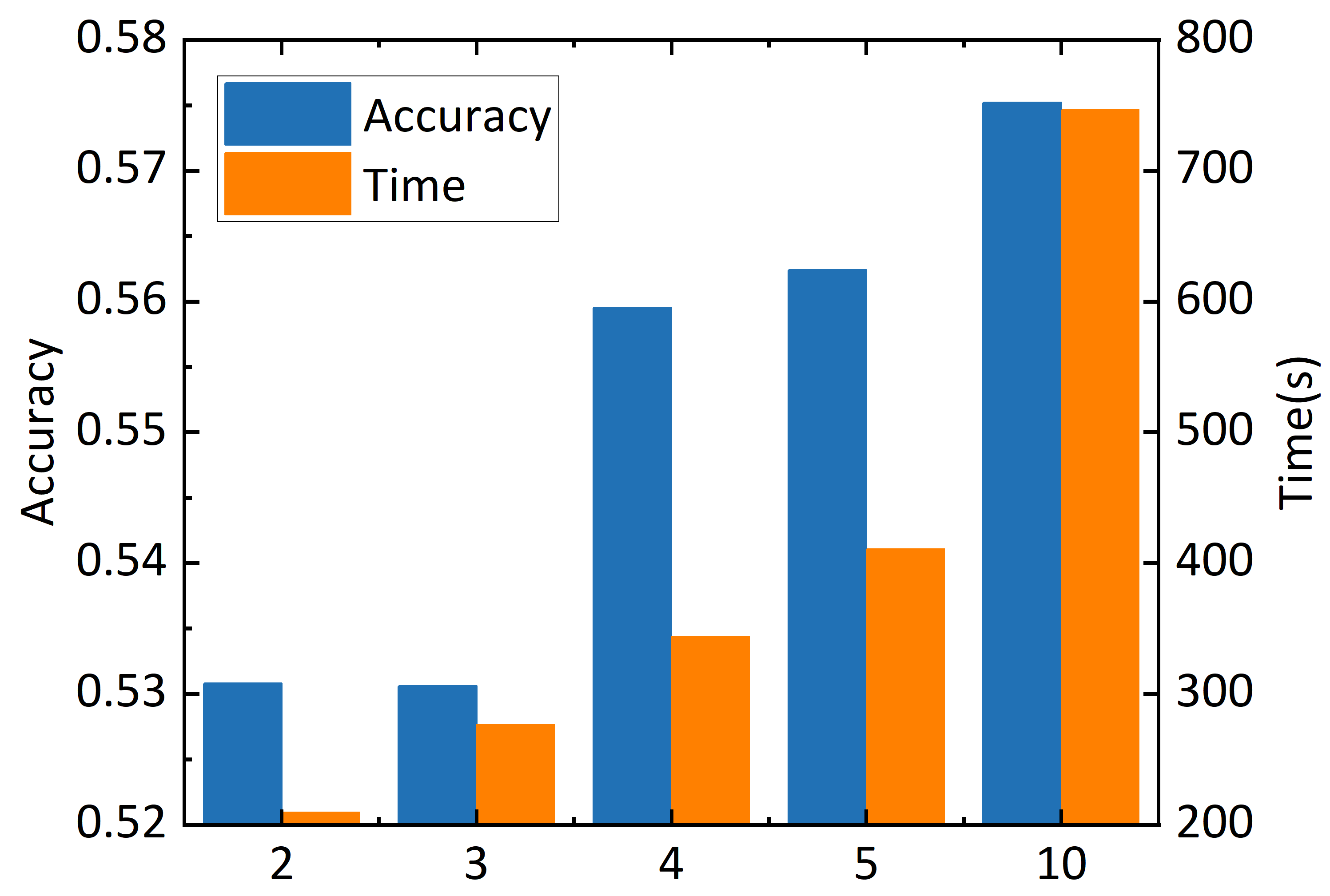}
        \vspace{-15pt}
        \caption{Effect of different $\tau_2$ values on CIFAR-10 dataset.}
        \vspace{-5pt}
        \label{fig:tau_2}
    \end{minipage}
    \hspace{0in}
    \centering
    \begin{minipage}[t]{0.275\textwidth}
        \centering
        \includegraphics[width=1.0\textwidth]{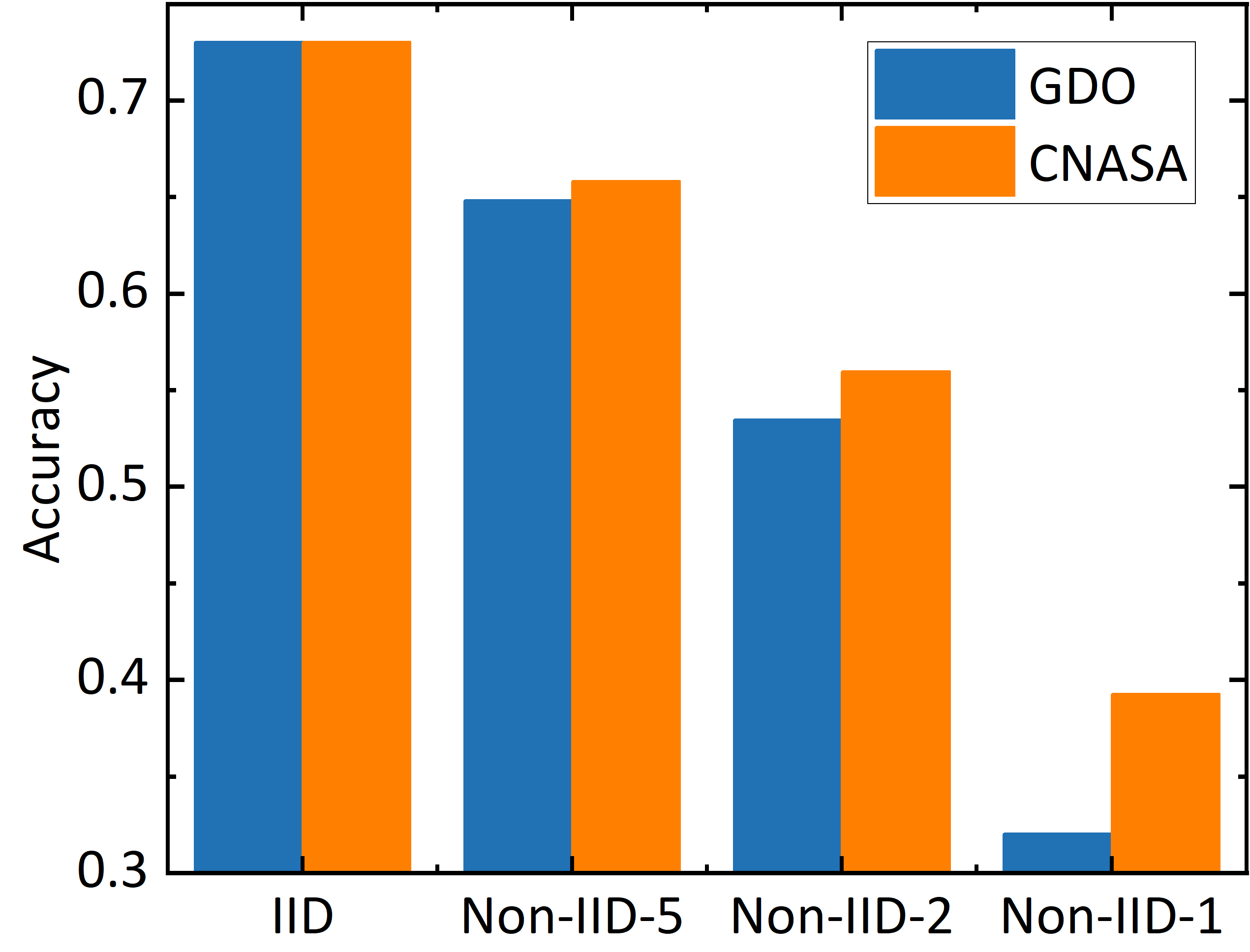}
        \vspace{-15pt}
        \caption{Test accuracy varies with different non-IID degrees.}
        \vspace{-5pt}
        \label{fig:Non-IID-degree}
    \end{minipage}
    \hspace{0.2in}
    \centering
    \begin{minipage}[t]{0.28\textwidth}
        \centering
        \includegraphics[width=1.0\textwidth]{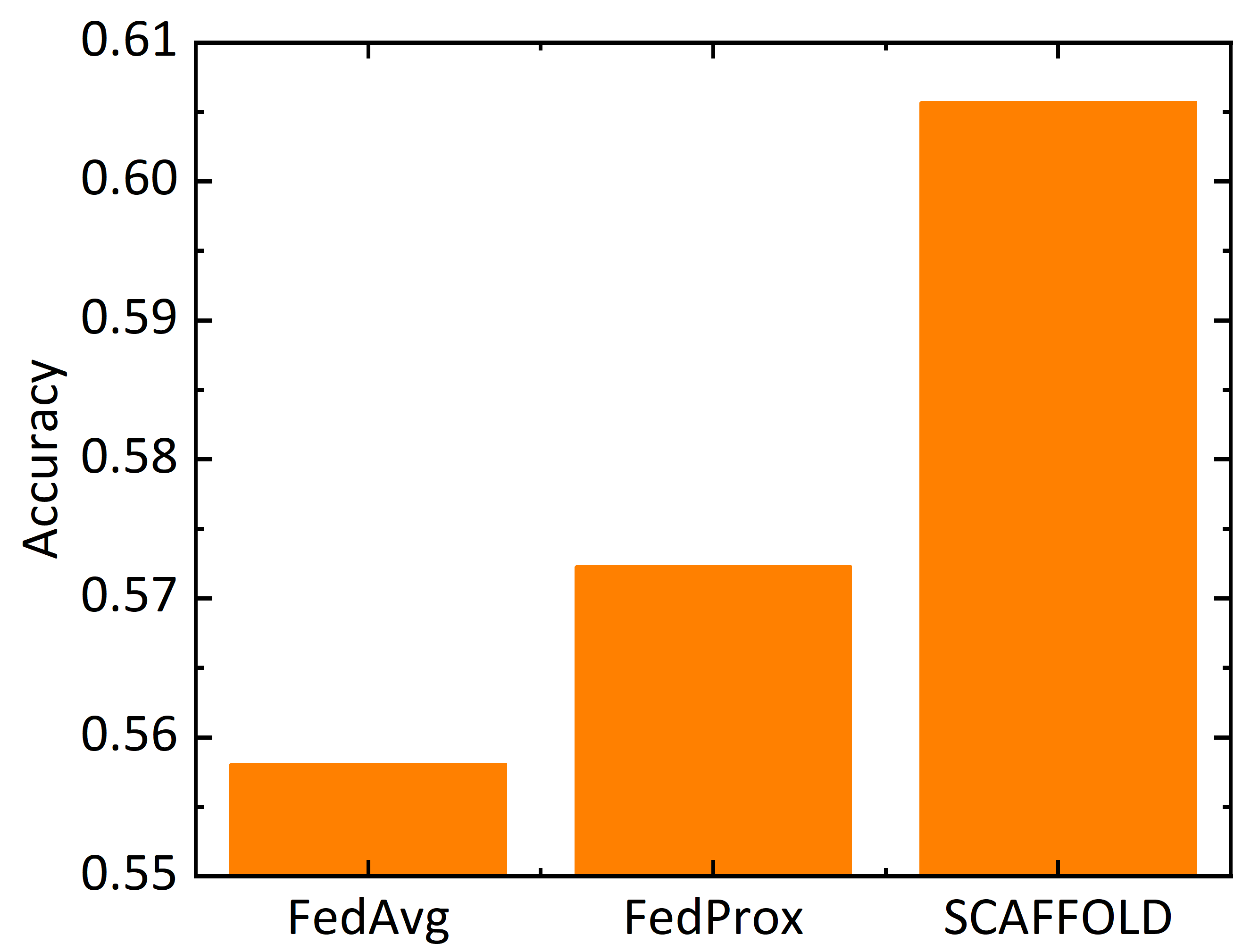}
        \vspace{-15pt}
        \caption{Test accuracy when using different FL schemes.}
        \vspace{-5pt}
        \label{fig:sca_prox}
    \end{minipage}
    \vspace{-10pt}
\end{figure*}
In this experiment, we only compare CNASA algorithm with GDO method, as CDO algorithm always requires too much time cost, although it can achieve the highest accuracy. As depicted in Fig.~\ref{fig:Non-IID-degree}.
both GDO and CNASA algorithms achieve the highest accuracy in IID case. As the degree of non-IID increases, the test accuracy of both the GDO method and our CNASA algorithm has decreased. However, the test accuracy of our CNASA algorithm is always higher than that of the GDO method, and as the degree of non-IID increases, the accuracy gap between our CNASA algorithm and the GDO method is getting bigger and bigger (e.g., in non-IID-1 case, the test accuracy of CNASA is 21\% higher than that of the GDO).
Therefore, our CNASA algorithm is more robust to the data heterogeneity. 
\subsubsection{Effect of different FL schemes in OBL}
For further model performance improvement, the OBL framework can adopt some state-of-the-art FL schemes (e.g., FedProx \cite{li2020federated} and SCAFFOLD \cite{karimireddy2020scaffold}) which alleviate data heterogeneity inherent in FL through elaborate aggregation mechanisms. As illustrated in Fig.~\ref{fig:sca_prox}, when using FedProx or SCAFFOLD for FL model aggregation, there exists $1.5-4.5\%$ accuracy improvement for Non-IID-2 case of the CIFAR-10 dataset. Hence, we can choose different FL schemes for our OBL framework based on diverse preferences and requirements of different FL learning tasks. 
\subsubsection{Performance of OBL under different numbers of IoRT devices, air nodes and satellites} 
\begin{figure*}[t!]
    \flushleft
    \begin{minipage}[t]{0.321\textwidth}
        \centering
        \includegraphics[width=1.0\textwidth]{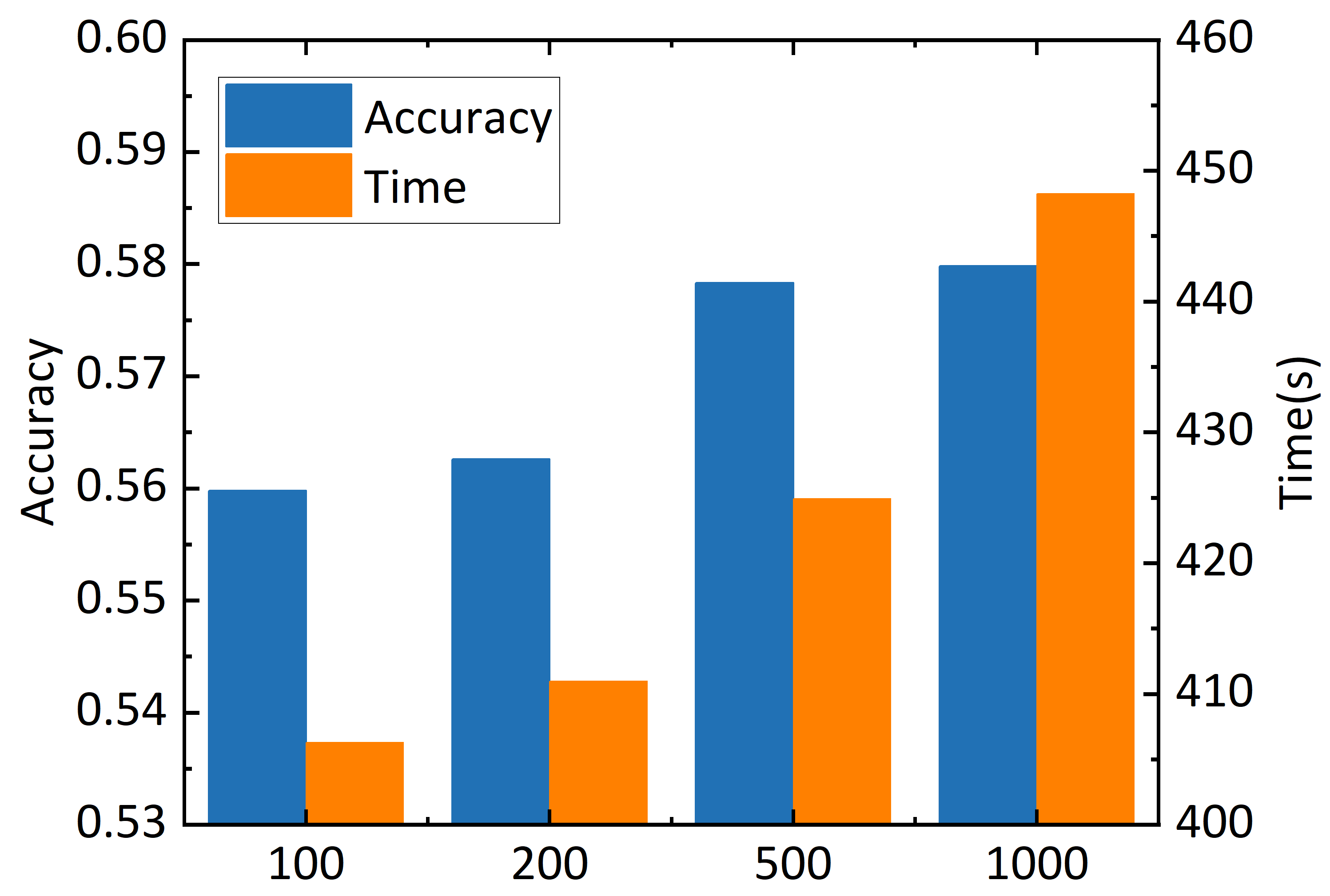}
        \vspace{-15pt}
        \caption{Effect of the number of IoRT devices on OBL.}
        \vspace{-5pt}
        \label{fig:device}
    \end{minipage}
    \hspace{0in}
    \centering
    \begin{minipage}[t]{0.321\textwidth}
        \centering
        \includegraphics[width=1.0\textwidth]{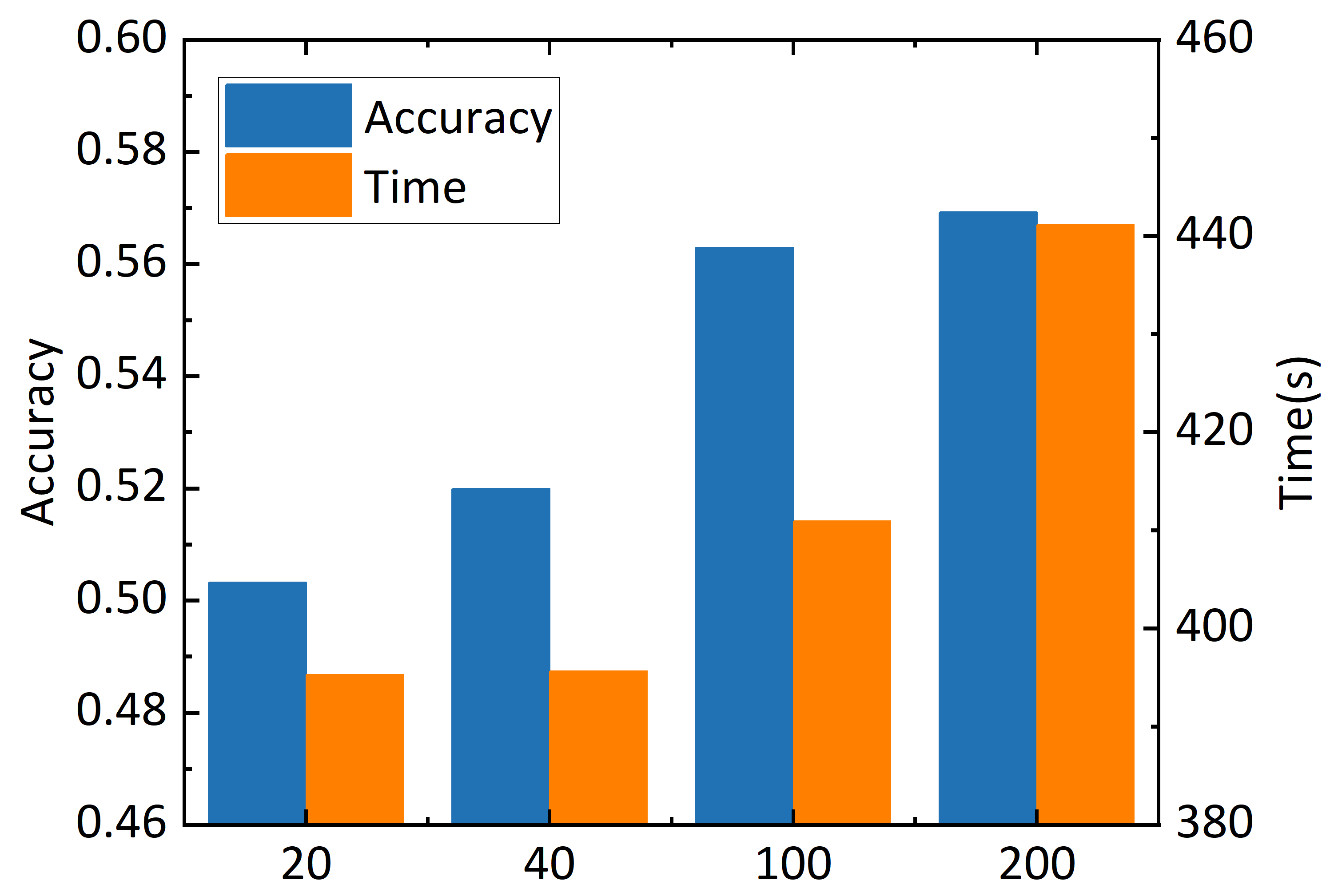}
        \vspace{-15pt}
        \caption{Effect of the number of air nodes on OBL.}
        \vspace{-5pt}
        \label{fig:air}
    \end{minipage}
    \hspace{0in}
    \centering
    \begin{minipage}[t]{0.321\textwidth}
        \centering
        \includegraphics[width=1.0\textwidth]{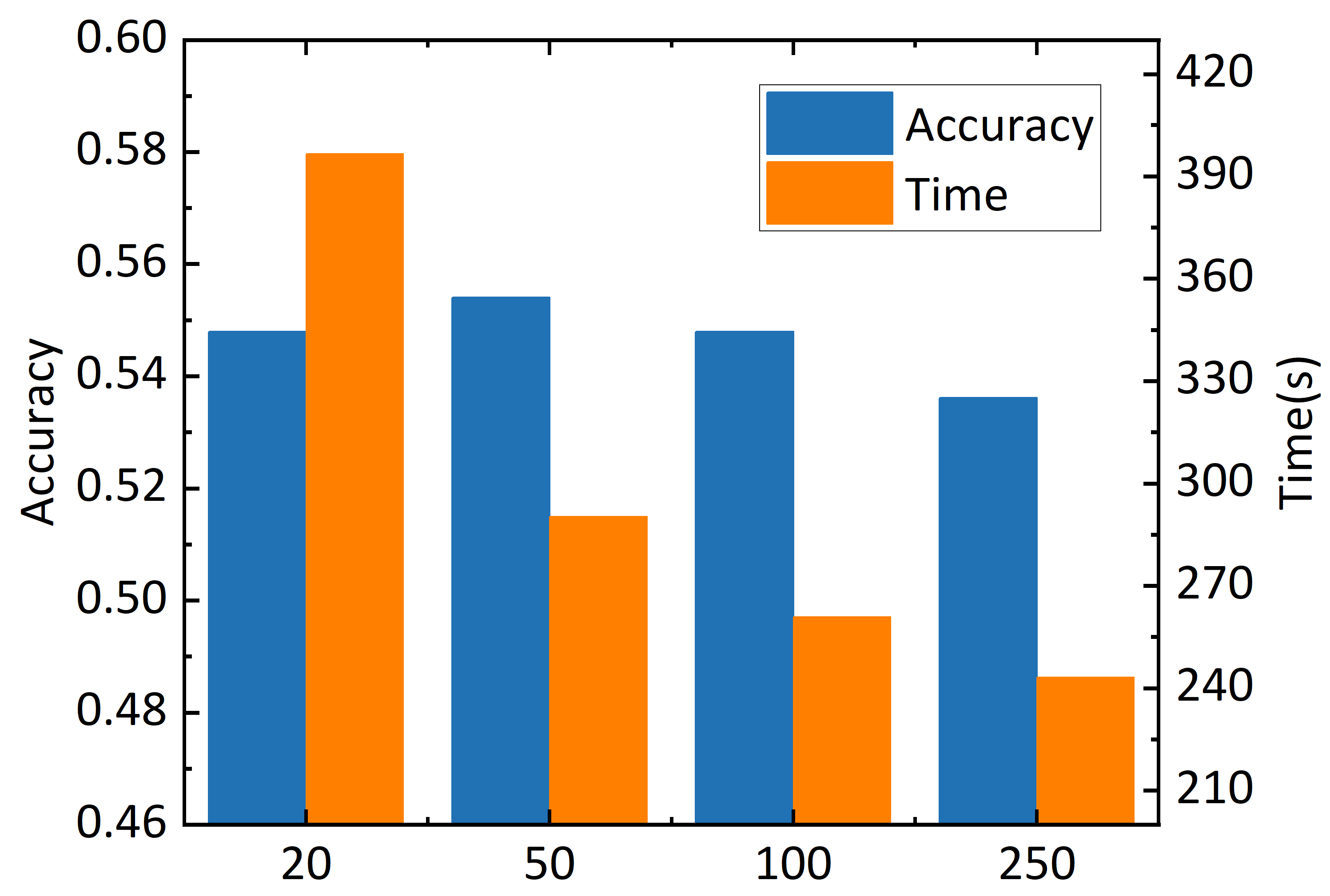}
        \vspace{-15pt}
        \caption{Effect of the number of satellites on OBL.}
        \vspace{-5pt}
        \label{fig:satellite}
    \end{minipage}
    \vspace{-10pt}
\end{figure*}

We further investigate the effect of the number of IoRT devices, air nodes and satellites in the SAGIN network on OBL algorithm with non-IID-2 case. As illustrated in Fig. \ref{fig:device}, the test accuracy of OBL continues to improve with the increase of the number of participating IoRT devices, as our OBL algorithm can leverage the rich data generated on more devices. In terms of time cost, as the number of IoRT devices increases, the bandwidth allocated to each device shrinks owing to parallel transmission, resulting in higher time cost.

In Fig. \ref{fig:air}, we can see that the test accuracy improves when increasing the number of air nodes. This is because that when the number of air nodes increases, each air node will cover a smaller number of IoRT devices, which can be seen as a finer-grained division of device data. Thus, our CNASA algorithm can conduct more elaborate air node-satellite assignment to mitigate the data heterogeneity and boost the model accuracy. Meanwhile, as the number of air nodes increases, so does the communication burden between satellite and air node, which will increase the time cost.

\begin{figure*}[t!]
    \flushright
    \begin{minipage}[t]{0.28\textwidth}
        \flushright
        \includegraphics[width=1.0\textwidth]{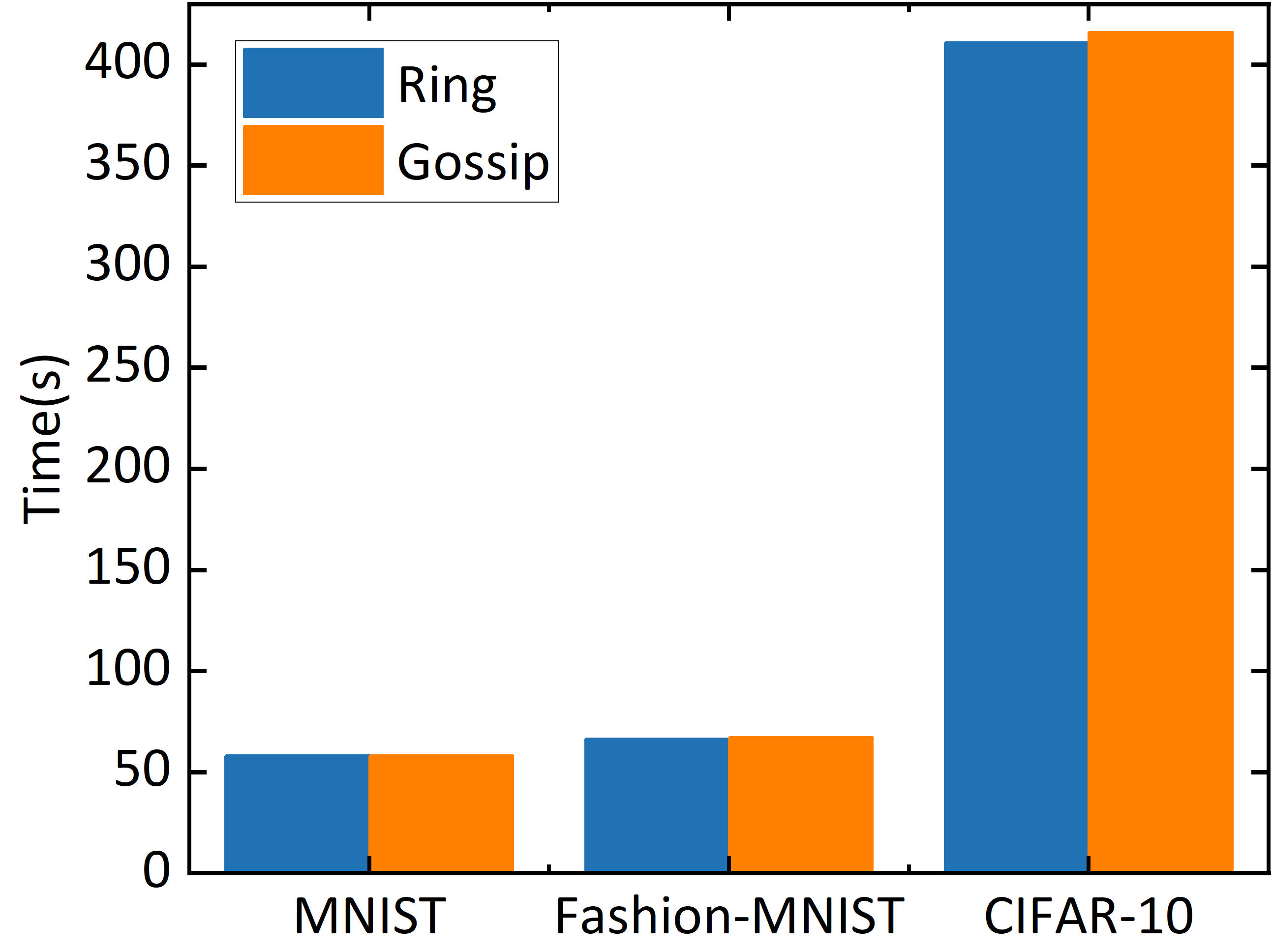}
        \vspace{-15pt}
        \caption{Time cost of Ring Allreduce algorithm and gossip protocol.}
        \vspace{-5pt}
        \label{fig:Ring_vs_gossip}
    \end{minipage}
    \hspace{0.15in}
    \centering
    \begin{minipage}[t]{0.333\textwidth}
        \centering
        \includegraphics[width=1.0\textwidth]{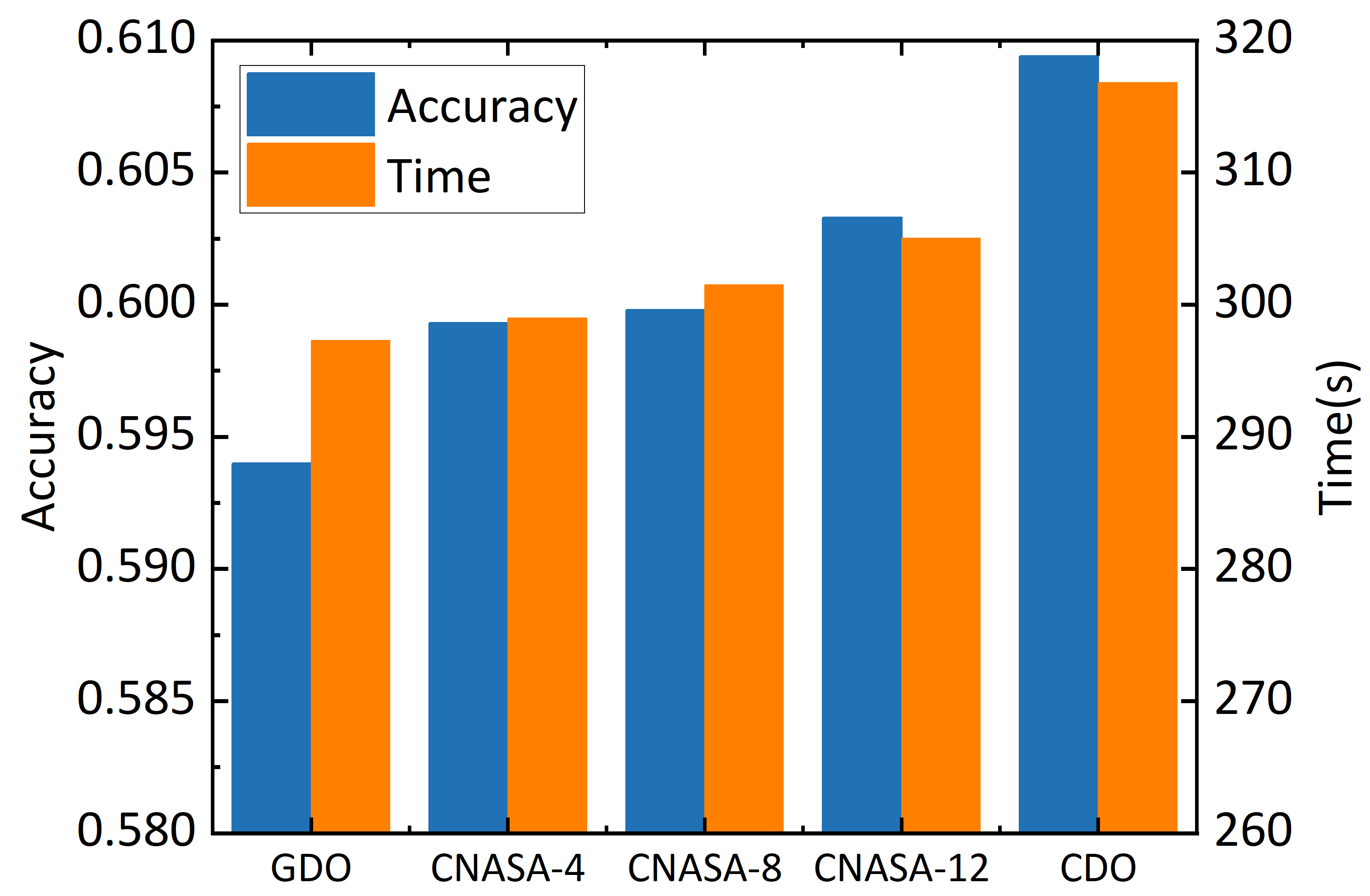}
        \vspace{-15pt}
        \caption{Test accuracy and time cost on the multi-orbit satellite network.}
        \vspace{-5pt}
        \label{fig:CIFAR10_acc_time_multi}
    \end{minipage}
    \hspace{0.005in}
    \centering
    \begin{minipage}[t]{0.327\textwidth}
        \centering
        \includegraphics[width=1.0\textwidth]{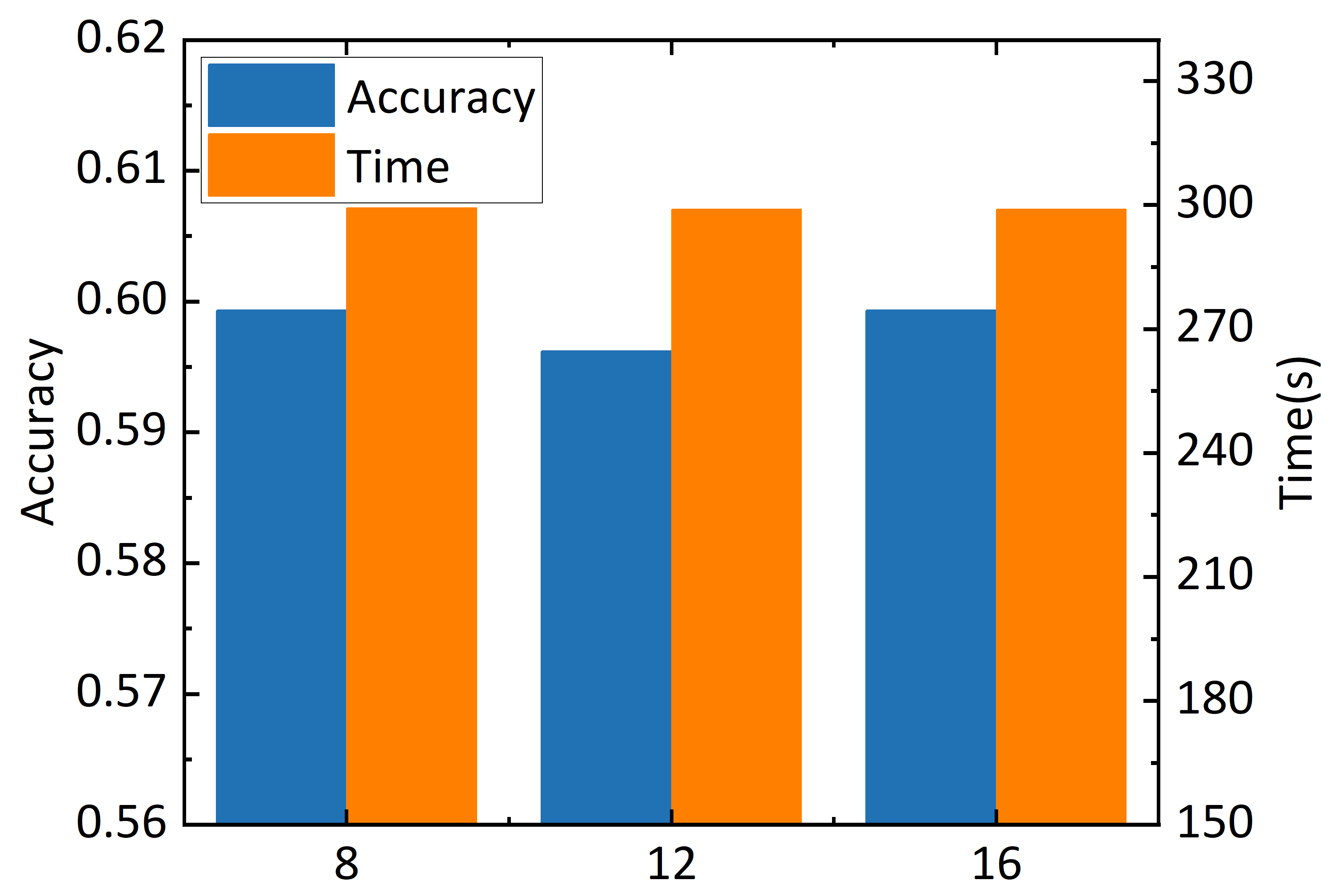}
        \vspace{-15pt}
        \caption{Effect of different numbers of orbits on CIFAR-10.}
        \vspace{-5pt}
        \label{fig:CIFAR10_acc_time_orbits}
    \end{minipage}
    \vspace{-10pt}
\end{figure*}

To evaluate the OBL performance under different satellite numbers, we assume that there are 500 IoRT devices and 500 air nodes in the SAGIN network. As depicted in Fig. \ref{fig:satellite}, the test accuracy remains at almost the same level with different number of satellites. The slight performance degradation at the case with 250 satellites is due to the serious non-IID data distribution at each satellite, as each satellite only in charge of two IoRT devices, indicating that the data samples for each satellite are from at most four classes (the data samples processed on each IoRT devices are from two classes in non-IID-2 case). Meanwhile, as the number of satellites grows, the communication load between satellites and air node decreases, allowing for greater bandwidth for each air node and hence lowering the time cost.
 
\subsubsection{Efficiency of Ring Allreduce algorithm}
To evaluate the efficiency of the Ring Allreduce algorithm in the inter-satellite model synchronization, we adopt the gossip protocol, a widely used aggregation algorithm in distributed systems, as the comparison algorithm. Specifically, in FL based on gossip protocol, each node sends its local model to its neighbor nodes, and the model is propagated node by node among the entire network. As illustrated in Fig.~\ref{fig:Ring_vs_gossip}, the time cost of the gossip protocol is higher than that of the Ring Allreduce algorithm. This is because the frequent model delivery between satellites in the gossip protocol incurs excessive communication overhead, while the Ring Allreduce algorithm distributes computation and communication evenly over all satellites by making good use of the ring topology formed by the satellites, thus shortening the model delivery times effectively.

\subsubsection{Evaluation on multi-orbit satellite network}

The performance of OBL on multi-orbit satellite network is evaluated on a SAGIN network with a LEO walker constellation as its space layer, as illustrated in Fig.~\ref{fig:Multi_orbit}. The walker constellation consists of 240 LEO satellites evenly distributed among 15 inclined orbital planes with an inclination of 85 degrees.
Due to limited GPU memory, we assume there are 480 air nodes and 480 IoRT devices in this SAGIN network.
For the locations of air nodes and IoRT devices, we assume that the entire earth's surface is divided by logical satellite positions filled by the nearest satellite, and each logical satellite location contains two air nodes and their respective IoRT devices, referring to the virtual node strategy in~\cite{8932318}.
The other network settings are the same as described in Table~\ref{table:network_settings}.
To mimic the non-IID scenario, We assign each IoRT device 5 classes of data samples in CIFAR-10 dataset.

We use CNASA-4, CNASA-8, and CNASA-12 to denote the extended CNASA algorithm with $N_{geo}=4$, $N_{geo}=8$, and $N_{geo}=12$, respectively.
As shown in Fig.~\ref{fig:CIFAR10_acc_time_multi}, the CDO method still obtains a final model with the highest test accuracy, i.e. 60.94\%, but its time cost is also the longest. Also, the GDO method still obtains the poorest final model, but its time cost is the shortest.
These results are consistent with those in the single-orbit satellite network.
Overall, our CNASA algorithm with $N_{geo}=12$ obtains a final model with a test accuracy of 60.33\%, which is only 0.6\% lower than that of the CDO method. Meanwhile, the time cost of CNASA algorithm with $N_{geo}=4$ is only 1.69s higher than that of the GDO method, which demonstrates that CNASA algorithm is efficient in the multi-orbit satellite network.

\subsubsection{Effect of the number of orbits}

We also investigate the effect of the number of satellite orbits on our CNASA algorithm.
Note that while the number of orbits has changed, the number of satellites remains the same, i.e., 240.
We fix $N_{geo} = 4$ in our CNASA algorithm.
As shown in Fig.~\ref{fig:CIFAR10_acc_time_orbits}, the number of orbits has little effect on either global model accuracy or time cost, reflecting that our CNASA algorithm is not sensitive to the number of orbits, that is, the CNASA algorithm can maintain good performance in various multi-orbit satellite networks.


%% file: Sections/Conclusion.tex
\section{Conclusion}
{\label{sec:Conclusion}}

In this paper, we propose OBL, a topology-aware federated learning framework for SAGIN, where the IoRT devices collaborate to train powerful models utilizing their local data with their privacy protected, and the nodes in the space layer and air layer serve as aggregation nodes.
To weaken the impact of non-IID data on IoRT devices and reduce the time cost of training a global model, we introduce the CNASA algorithm for OBL, which takes both the geographic distance and class distribution into account.
To adapt the OBL framework to the multi-orbit satellite network, we re-design the determination of each satellite's coverage area, the inter-satellite model aggregation mechanism, and the CNASA algorithm.
We also theoretically analyze the convergence of the OBL framework.
Experiment results demonstrate that models trained using our proposed OBL framework can achieve satisfactory accuracy while reducing the time cost in both single-orbit and multi-orbit satellite networks, which can also be applied to tasks with diverse requirements for accuracy and time cost.
Additionally, the OBL framework is robust to the heterogeneity of local data and can efficiently aggregate the global model.